\documentclass[11pt]{article} 
\usepackage{color}
\usepackage[utf8]{inputenc} 
\usepackage[T1]{fontenc}
\usepackage{currvita}
\usepackage{eurosym}
\usepackage{hyperref}
\usepackage{esvect}
\usepackage{amsmath}
\usepackage{caption}

\usepackage{geometry} 
\usepackage{graphicx} 

\usepackage{booktabs} 
\usepackage{array} 
\usepackage{paralist} 
\usepackage{verbatim} 
\usepackage{subfig} 

\usepackage{fancyhdr} 
\usepackage{sectsty}
\allsectionsfont{\sffamily\mdseries\upshape} 

\usepackage[nottoc,notlof,notlot]{tocbibind} 
\usepackage[titles,subfigure]{tocloft} 

\title{Microeconomic Structure determines \\
Macroeconomic Dynamics.\\
Aoki defeats the Representative Agent}
\author{Sorin Solomon and Nata\v{s}a Golo, \\
The Racah Institute of Physics, \\
Hebrew University of Jerusalem}

\begin{document}

\maketitle
\begin{abstract}
Masanao Aoki developed a new methodology for a basic problem of economics:  deducing rigorously the macroeconomic dynamics as emerging from the interactions of many individual agents.
This includes deduction of the fractal / intermittent fluctuations of macroeconomic quantities from the granularity of the mezo-economic collective objects (large individual wealth, highly productive geographical locations, emergent technologies, emergent economic sectors) in which the micro-economic agents self-organize. 

In particular, we present some theoretical predictions, which also met extensive validation from empirical data in a wide range of systems:
\begin{itemize}
\item The fractal Levy exponent of the stock  market index fluctuations equals the Pareto exponent of the investors wealth distribution. The origin of the macroeconomic dynamics is therefore found in the granularity induced by the wealth / capital of the wealthiest investors.  
\item Economic cycles consist of a Schumpeter 'creative destruction' pattern whereby the maxima are
 cusp-shaped while the minima are smooth. In between the cusps, the cycle consists of the sum of 2 'crossing exponentials': one decaying and the other increasing. 
\end{itemize}
 
This unification within the same theoretical framework of 
short term market fluctuations and long term economic cycles offers the perspective of a genuine
conceptual synthesis  between micro- and macroeconomics.
 Joining another giant of contemporary science - Phil Anderson \cite{Anderson 1972} - Aoki emphasized the role of rare, large fluctuations in the emergence of macroeconomic phenomena out of microscopic interactions and in particular their non self-averaging, in the language of statistical
physics. In this light, we present a simple stochastic multi-sector growth model.

\end{abstract}

\section{Introduction}
\label{sec:intro}
I first met Professor Aoki some 15 years ago when the application of physics inspired methods was a rare appearance at economics conferences. Yet already then Masanao had an impressive amount of work applying statistical mechanics techniques to economic issues such as growth, employment, aggregation. Since then, we kept meeting almost yearly at workshops, schools and conferences. Beyond the technical interest in his work, I felt great hope that Masanao scientific vision and personality might help us achieve this fusion between the economic facts and ideas and between the physics techniques and Popperian ethos \cite{Popper 1976}.  Yet, the physical oceans separating us prevented what a posteriori feels like a great occasion of merging more intimately our ideas and methods. We take the present volume as a belated occasion to draw the lines along which such a synthesis or at least dialogue can take place.  We will first list some of the converging interests, ideas and subjects and then will concentrate on a few more specific results. 
\section{Solvable agent based models of Schumpeter's creative destruction }
\label{sec:growth}
Following Schumpeter \cite{Schumpeter 1936}, Aoki and Yoshikawa attributed growth to the endogenous emergence of novelty.  The introduction of new goods, the opening of new markets, the financial innovations do influence the firms of the real sector heterogeneously. After such an event (/shock), many big firms start to shrink and some small firms start to grow. Moreover, firms in previously dominating sectors disappear exponentially, while new sectors and firms capable of exploiting the new situation are established in growing numbers (Fig. \ref{fig:crossing}). Schumpeter and the later evolutionary economists have (sometimes with reserve) likened this to biological evolution. Aoki and Yoshikawa, following Montroll \cite{Montroll 1978} emphasized another common factor influencing economic and biological phenomena: the Malthus-Verhulst logistic equation. Aoki considered very early \cite{Aoki 1968} a simpler version of such a dynamics which in fact is mathematically similar to the famous  AK model \cite{Romer 1986} Eq. \ref{eq:Eq25}.

 In the present paper we elaborate on the connection between the theoretical predictions of such an approach and the empirical observations on the economic effects of technological, financial and political changes.  
Moreover we explain, fit and predict the fractal / intermittent macroeconomic  fluctuations 
\footnote{In this paper we use for fractal fluctuations the definition of Mandelbrot for Lévy flights \cite{Mandelbrot 1982}: a random walk in which the step-lengths have a power law (Pareto) probability distribution. 
For intermittency we use the definition of Zel'dovich \cite{Zeldovich 1987}:"Some specific structures in which a growing quantity reaches record high values typically arise
for instabilities in random media. Despite the rarity of these concentrations, they dominate the
integral characteristics of the growing quantity (the mean value, the mean square value, etc.).
The appearance of such structures is called 'intermittency'."
Thus we reserve the term 'intermittent' for random processes totally dominated by their rarest events
while for processes that are merely outside the basin of attraction of the normal distribution we use the term fractal.}
as a result of the granular / multi-scale structure of the collective objects  (capital accumulations, economic sectors, geographical regions with outstanding productivity) that aggregate the microeconomic elements (money, investors, workers, technological know-how).

By doing so we transcend \cite{Kirman 1992} the 'Representative agent' approach giving macroeconomics a microeconomic foundation that is able to provide, in coherent conceptual framework, theoretical and numerical predictions in agreement with the empirical facts. 
Such multi-agent non-equilibrium models are not the bread and butter of the neo-classical economics mainstream. In the past many economists \cite{von Hayek 1937} deemed it impossible to deduce from first principles the collective dynamics underlying the macroeconomic properties. But Masanao was not afraid to face these difficulties head-on. He used the master (Chapman-Kolmogorov) equation as a powerful method for the correct aggregation of 'microeconomic acts' into 'macroeconomic behavior' erasing thereby de facto the barrier between micro- and macro–economics \cite{Aoki 2004}.  By applying exact methods he showed that in economic systems the fluctuations are not vanishing even when the number of components is taken to infinity:  the statistical samples are not self-averaging and lead to macroscopic (fractal, intermittent) fluctuations. In this context, he found rigorous basis for the emergence of power laws in economic statistical distributions.
  Masanao used the master (Fokker-Planck) equations also in order to characterize the clustering effects and the space-time aggregation processes responsible for macroeconomic fluctuations.  Such effects are indeed crucial for the economic growth. 

Aoki's conclusions converge with the results obtained by physics techniques \cite{Malcai 1999} which predicted theoretically and validated empirically the equality between the Pareto wealth distribution exponent and the fractal exponent characterizing the fluctuations of the stock markets and growth rates Fig. \ref{fig:Pareto} and \cite{Solomon 2003}.This was an important step in transforming Economics into a Popperian science: the theoretical prediction connecting the individual wealth granularity to the fluctuations of aggregated economic and financial indexes has been very precisely validated by the empirical measurements \cite{Klass 2006}  \cite{Klass 2007} \cite{Gabaix 2011}
Such causal models which predict connections between structural (granularity) and dynamical (fractal / intermittent fluctuations) measurable properties of the economic systems provide a significant advancement beyond the mathematical identities connecting static variables provided by the neo-classical equilibrium assumptions.

Aoki’s work connotes with Schumpeter proposal \cite{Schumpeter 1936} of innovation as the intrinsic fundamental feature of capitalism. According to Schumpeter, in order to achieve progress after an innovation, it is not sufficient to establish and develop the new firms and sectors applying it for production: it is equally important for the production means and organization related to the old technology to disappear (cf. Fig. \ref{fig:crossing} below). Thus Schumpeter coined the term 'creative destruction' which makes bankruptcies, crisis and unemployment part of the economic progress. It connects the booms and crises in the global economic indices to the appearance, disappearance, shrinking and growing of collective economic objects (firms, investor herds, large wealth accumulations, sectors, technologies, regions) and their production.

Given the limitations in the mathematical methods available for multi-agent systems at the time, Schumpeter’s model was not formally and quantitatively tractable.  As expressed mathematically by the non-self-averaging properties emphasized by Aoki, an economic system with many interactive agents and multi-scale granularity is exceedingly difficult to predict. Yet the recognition of its importance for describing economic cycles continued to grow \cite{Dosi 2012}. 
Masanao’s vision and methods may find one of their most impactful applications in this direction.  Following the appearance of a new technology or new production conditions the economic system may undergo extensive changes: following such a shock, some sectors which might not even existed before, emerge, grow and their companies multiply. By contrast, some other sectors which were dominating before the change become not sustainable, they shrink and their companies disappear. This leads to the emergence of very interesting adaptive collective objects (new sectors, new technologies, new 'Silicon Valleys', new 'Start-Up Nations') \cite{Yaari 2008}. 
In turn, the growth / shrinking / appearance / failure of the individual firms, sectors, technologies constitute the very stuff from which the global intermittent macroeconomic dynamics is made of. They are in the same time an embodiment of the Schumpeter creative destruction scenario: the old economy is shrinking and innovative sectors appear and develop adaptively \cite{Challet 2009}, \cite{Dover 2009}. 

We will see that this picture has very precise implications on the short (Fig. \ref{fig:Pareto})
 and long (Figs. \ref{fig:crossing} , \ref{fig:double})
 term fluctuations of the economic indices. 
These implications were confirmed with great precision by the empirical data Figs. 
\ref{fig:Pareto}, \ref{fig:Balkan}, \ref{fig:Poland}.

\section{ Proliferating agents and logistic / autocatalytic systems }
\label{sec:logistic}
The crucial property that fuels the propagation of microeconomic events to systemic changes is 'autocatalicity'. By the term 'autocatalytic' we mean here a process or rather a quantity whose time variation is proportional to itself as expressed mathematically by Eqs. \ref{eq:malthus}-\ref{eq:Malthus solution} and \ref{eq:Eq25}-\ref{eq:MatrixSolution}.  The origin of the concept of autocatalytic processes as the mechanism of systemic growth can be traced back 200 years to Malthus  \cite{Malthus 1798}. Malthus ideas have been developed extensively by later workers. Aoki cites Montroll \cite{Montroll 1978} as attributing Malthus-like dynamics to virtually all social systems. Malthus applied it initially to the population growth
 but we shall see that the idea is relevant to economic growth and many other phenomena \cite{Shiller 2000}. 

In modern agent-based language, Malthus' assumptions can be translated at the individual agent 'microscopic' scale as follows: 
\begin{itemize}
\item in the presence of sufficient resource units $a$ any (female) individual (animal / human) $k$ can  generate with some probability rate per unit time $s$ another (female) animal / human:
\begin{equation}
k+a \rightarrow k+k+a; \; \; s,
\label{eq:s}
\end{equation}
\item each such individual has a death probability rate $\delta$ per unit time:   
\begin{equation}
k \rightarrow nothing; \; \; \delta.
\label{eq:delta}
\end{equation}
\item each $k$ and $a$ can diffuse by jumping between neighboring $x$ locations with probabilities per unit time $D_K$ and respectively $D_A$.
\end{itemize}

Malthus implicitly assumed that the space $(x)$ and time $(t)$ density $(A(x,t))$ of the $a$ agents is constant, i.e. $A(x,t)=A$. This seemed as a good assumption at the time. For instance, this is the macroscopic asymptotic equilibrium state if one assumes that the $a$’s  diffuse randomly cf. Eq. \ref{eq:Laplace}. We will see later that this approximation that neglects the microscopic random fluctuations intrinsic to the discrete character of the $a$ individuals, is highly dangerous and misses most of the salient effects. By neglecting the space dependence of the $a$ and $k$ agents densities $A(x,t)$ and $K(x,t)$
\footnote{We shall call occasionally  $A(x,t)$ and $K(x,t)$ 'densities' when discussing them in the macro- approximation where  $A(x,t)$ and $K(x,t)$  are considered as functions defined on a space where the locations  $x$ are real numbers. However one should not forget that the main point of this paper is that the macro / continuum approximation has severe limitations and the correct and binding formulation is the discrete one, where $A(x,t)$ and $K(x,t)$ are just the number of agents $a$ and $k$ at the discrete location $x$ at time $t$.}
, Malthus deduced an ordinary differential equation governing the time evolution of $K(t)$, the total number of  $k$'s in terms of $A$, the reproduction rate $s$ and the death rate $\delta$. 
\footnote{The notations $K$, $s$, $A$, $\delta$ are not the standard ones in the Malthus context. We use them here to facilitate the contact with the economist readership: with the present notations, Eq. \ref{eq:malthus} becomes identical with the familiar AK model Eq. \ref{eq:Eq25}  \cite{Aghion 2008}. }:
\begin{equation}
\frac{dK}{dt}=(sA-\delta)K(t)=gK(t),
\label{eq:malthus}
\end{equation}
where one defined the growth rate $g$ by:
\begin{equation}
g=sA-\delta
\label{eq:growth rate}
\end{equation}
Eq. \ref{eq:malthus} has the exponential solution
\footnote{ A few decades after Malthus, Verhulst \cite{Verhulst 1838} included in Eq. \ref{eq:malthus} a non-linear term $-K^2$ corresponding to  $k$'s competition, confrontation, limited resources and called the new equation \emph{'the logistic equation'}.  In terms of the agents the competition between $k$’s is expressed by the reaction:
$ k+k \rightarrow k;  \; c. $
The Verhulst non-linear term has the effect of diminishing the $K(t)$ growth, ${dK}/{dt}$, as $K(t)$  increases and in fact eventually leads to the saturation of the $K$ population growth.  Thus, the solution of the logistic equation starts increasing in time with the same exponential growth rate  $g$  as the Malthus solution Eq. \ref{eq:Malthus solution}, but eventually it curbs down and saturates. In the current discussion we will not discuss the effects of the Verhulst term.  }
:
\begin{equation}
K(t)=K(0)  e^{gt}
\label{eq:Malthus solution}
\end{equation}
i.e. exponential growth for 
\begin{equation}
g>0
\label{eq:glarge}
\end{equation}
and exponential decay for 
\begin{equation}
g<0 .
\label{eq:gsmall}
\end{equation}

For two hundred years it was believed that the differential equations of the type Eq. \ref{eq:malthus} can faithfully represent the evolution of the total amount $K(t)$ of $k$ self-reproducing agents 
\footnote{ We will use occasionally the integral $\int_x  $ and differential $\Delta$ notations when 
discussing the continuum approximation. However, one should remember that it is one of the 
main claims of the present paper that this approximation fails in crucial ways and that the 
correct formulation is the discrete one. In particular using discrete sum instead of the integral. }:   
\begin{equation}
K (t) = \int_{x} K(x,t) dx = \sum_x K(x,t)
\label{eq:Klarge}
\end{equation}
in the presence of an average amount of resources
\begin{equation}
A (t) = < A(x,t) >_x 
\label{eq:Alarge}
\end{equation}
(recall that since $a$ are not created or destroyed $A(t)=A(0)$ is constant in time).

In fact, even when one starts with a macroscopically inhomogenous $A(x,t)$, the diffusion of the individual $a$'s, expressed in the continuum as:
\begin{equation}
{\partial A (x,t)} / { \partial t} = D_A \Delta A(x,t)
\label{eq:Laplace}
\end{equation}
causes A(x,t) to converge rather fast  to a spatially and temporally constant $A(x,t)= < A(x,0)>_x =  A(0) $.

So once one neglects the microscopic fluctuations of $A(x,t)$ and considers it a continuous differentiable function,  one is forced unavoidably into the Eqs. \ref{eq:malthus} and \ref{eq:Malthus solution}.
However its  has been shown that the very assumption that the system of discrete agents Eqs. \ref{eq:s}, \ref{eq:delta} can be faithfully represented in terms of continuous differentiable functions  $A(x,t)$ and $K(x,t)$ turns out to be false. If the growth factor is different for different locations and parts of the system, then the different exponentials of the type Eq. \ref{eq:Malthus solution} at different locations lead to functions $K(x,t)$  which are too singular to respect (partial) differential equations. In particular, the naïve scalar ordinary differential equation Eq. \ref{eq:malthus} does not apply. 

We refer the reader to the original literature  \cite{Shnerb 2000}, \cite{Kesten 2002}, \cite{Louzoun 2007}
 for the rigorous proofs and present below only  heuristic explanations and discussions of it. 
The most striking result obtained in these papers is that in large enough spaces with $d \le 2$ dimensions the $k$ population $K(t)$ always grows irrespective of the values of $s$, $\delta$, $A$, $D_A$, $D_K$.
The proof goes along the following points:
\begin{itemize}
\item one assumes that each agent $a$ has an equal probability to be originally situated 
at any of the locations $x$ in space.
This means that the $a$'s are distributed by a Poisson distribution  Eq. \ref{eq:Poisson}.
\item Random walks in $d \le 2$ dimensions are recurrent. 
I.e. as the duration of the random walk is taken to infinity, each agent $a$ returns an infinite number of times to its original position.
\item during those visits 'home' to its site of origin $x$, each $a$ contributes an exponential factor to the $K(x,t)$ expectation.
\item So, each $a$ originating at $x$ has a multiplicative contribution to $K(x,t)$ expectation that increases exponentially in time.
\item Thus, if at the beginning there are enough $a$ agents $A(x,0)$ on a site $x$ then the product of their exponential growth contributions will be able to defeat / compensate any exponential decay with fixed negative rate originating in the $k$’s deaths (due to $\delta$) and $k$ migration (due to $D_K$).
\item Even if the above scenario requires a very large initial number $A(x,0)$ of $a$'s to be originally at the site $x$: $A(x, 0) >> A_{average}$ still this has a finite (even if exceedingly small)  probability $P(A(x,0) )$ .
In fact one can compute it using the Poisson formula Eq. \ref{eq:Poisson}.
\item If the volume (number of sites $x$) of the system is $V >> 1/P(A(x,0) )$, then the expected number of sites $x$ where one has initially a number of $a$ agents $A(x,0)$ or more is larger than 1.
\item In those points, the expected value of $K(x,t)$ grows exponentially.
\item Thus even if in all the other points $K(y,t)$ collapses, the total $K (t)$ grows exponentially on the account of those special / singular points. 
\item The expected growth of $K(t)$ is an exponential of time with a coefficient of $t$ in the exponent which is dictated by the largest $A(x,0)_{max}$ in the system. In particular for an infinite system , the expectation of $K(t)$ grows faster than any exponential \cite{Kesten 2002}.
\end{itemize}

Thus it comes natural that in the next section we first describe the case in which the $a$'s do not leave the original site at all ($D_A =0$) and then to extend the analysis to the case $D_A > 0$ while taking into account the occasional returns of $a$ to their original site.

\section{Anomalous resilience of autocatalytic multi-agents systems}
\label{sec:resilience}

If we look at the system Eqs. \ref{eq:s}, \ref{eq:delta} microscopically, from the individual agents point of view, even if the position of each of the $a$'s is chosen from a spatially uniform probability distribution, the result will be that each realization of the system will respect a Poisson distribution. In particular each finite (but arbitrarily large) value of $A(x,t)$ at any location $x$, has a finite (though possibly extremely small) probability to be realized.   Ignoring the rarest events where $A(x,t)$ is largest leads to the naive (and too stringent) criterion, Eqs. \ref{eq:glarge},  \ref{eq:growth rate}, for the survival and growth of $K (t)$:
\begin{equation}
A>\delta /s
\label{eq:naive}
\end{equation}
This is of course a too strong condition: for the total number $K(t)$ of agents $k$ to grow it is not necessary that $K(x,t)$ will grow at the majority of sites $x$ and times $t$.
The growth of $K(x,t)$ in a (non-zero-measure) subset $S(t)$ of the $x$ space is sufficient to insure the eventual growth of $K(t)$ as a whole .

A hint of how it happens is to realize that because of the convexity of the exponential function, the average of the exponents is larger then the exponent of the average. 
In our case, the average growth factor $\left< e^{g(x,t)} \right> _{x}$ is always larger than the exponent of the average growth rate $<g(x,t)>_{x}$:
\begin{equation}
\left < e^{g(x,t)} \right> _{x} > e^{<g(x,t)>_{x}}=  e^g
\label{eq:convexity}
\end{equation}

Thus even if the average of the growth rate $g$  is negative (Eq. \ref{eq:gsmall}), and the solution Eq. \ref{eq:Malthus solution} of the naive Eq. \ref{eq:malthus} is decaying exponentially, in the actual stochastic discrete system defined by Eqs. \ref{eq:s} and \ref{eq:delta} the microscopic granularity of $A(x,t)$ is sufficient to insure that in a very wide range of conditions the total number of $k$'s in the system $K(t)$ increases.

\subsection{The case of non-diffusive $a$'s: $D_A$ =0 }
\label{sec:non-diffusive}

For a first familiarization with the arguments, let us momentarily neglect  the diffusion of the $a$ agents and take $D_A=0$.  Assume each agent $a$ has an equal probability to be located at any of the locations in the $x$ space. 
Such a probability distribution was of course studied extensively under the name of Poisson distribution.

While at the macroscopic scales the Poisson distribution is as uniform as it can possibly be,  the microscopic granularity allows for arbitrarily large local deviations of $A(x,t)$ from the average $ A$.
In fact the probability at any site $x$ for the number of agents $a$ to be $A(x,t)$ is given by the Poisson formula:
\begin{equation}
P(A(x,t)) =  { A}^{A(x,t)}  e^{- A} / A(x,t) {!} 
\label{eq:Poisson}
\end{equation}
Thus, for any $x$, $A, \delta, s$ there is a non-vanishing (though possibly extremely small) probability that  $A(x,t) >\delta /s$.  Thus, for a large enough $x$ space,  with a number of locations $ N > 1/P(A(x,t))$ the expected number of sites $x$ with an $a$ occupancy $A(x,t)$ exceeds 1. In general, this means that in the $x$ space there exists a set $S(t)$
\footnote{Actually $S(t)$ does not change in time for the case $D_A=0$. But for the next case $D_A \ne 0$ the changes in $S(t)$ will be crucial.}
 of non-vanishing measure for which the local growth rate $g(x,t)$ is positive:
\begin{equation}
g(x,t)=sA(x,t)-\delta > 0; \ for \  x \in S(t)
\label{eq:growth g}
\end{equation}

 This is enough to insure that
\begin{equation}
K(t)=K(0) \int_x  e^{\int_t {g(x,t)dt}} dx  \rightarrow \infty.
\label{eq:granularity}
\end{equation} 

Eq. \ref{eq:granularity} holds  because even if for the rest of the $x$ locations the integral vanishes, on the set $S(t)$ the integral diverges:  
\begin{equation}
K(x,t) = e^{g(x,t)t}  ; \  for  \  x \in S(t)
\label{eq:growth exponential}
\end{equation}
(because $g(x,t)$ as well as $A(x,t)$ and $S(t)$ are actually constant in time if $D_A=0$).
Consequently 
\begin{equation}
K(t)  \sim  \sum_{x\in S(t)} e^{t {g(x,t)} } >   e^{t g_{max}} \rightarrow \infty.
\label{eq:exp max}
\end{equation} 
where $g_{max}$ is the largest among the values of $g(x,t)$ in the entire system.
Using the inequality \ref{eq:exp max}, Kesten \cite{Kesten 2002} proved that for an infinitely large system (infinite number of $x$ sites) $K(t)$, the number of $k$'s, increases even {\it faster} than any exponential.

Thus, the very rare microscopic events where $A(x,t)>\delta/s$ constitute singular growth centers where auto-catalytic amplification of the $k$'s density $K(x,t)$ takes place
and thus leads to the emergence in their neighborhood of large growing  $k$-collectives/ 'herds' / 'islands' / 'mountains' with exponentially growing $K(x,t)$. The caveat is that for $D_A\ne 0$  the $a$’s configuration changes continuously due to their diffusion and so does the growth set $S(t)$. Thus the crucial question for the survival and growth of $K(t)$ is whether the $k$ 'herds' / 'islands' / 'mountains' are able to follow the fluctuations, changes, appearance, disappearance  and moves of the  $A(x,t)>\delta/s$ regions $S(t)$.

\subsection{The case of diffusive $a$'s : $D_A \ne 0$}
\label{sec:Polyaresilience}

The question of the previous paragraph can be decided rigorously, analytically and quantitatively by treating the spatially extended agent based model defined by Eqs. \ref{eq:s}, \ref{eq:delta} using the combinatorial techniques of Masanao or by field theory renormalization group techniques \cite{Shnerb 2000}, or by branching random walk techniques \cite{Kesten 2002}.
The answer is that in the general case  where one considers the $a$'s diffusion $D_A >0$, the actual condition for $K(t)$ not to vanish as $t \rightarrow \infty$ is:
\begin{equation}
s/D_A>1-Pol_d,
\label{eq:Polya}
\end{equation}
where  $Pol_d$ is the Polya constant in the space with $d$ dimensions where the $a$’s diffuse. 
By definition the Polya constant $Pol_d$  is the probability that given enough time a randomly walking $a$ would return to its initial position. 
Since for $d \le 2$ space dimensions $Pol_d =1$, for a large enough system $K(t)$ never decays to $0$ because cf. Eq. \ref{eq:Polya} the condition becomes $s/D_A > 0$  which always holds for $s>0$. 

Note how different is the survival / growth condition Eq. \ref{eq:Polya} from the naive criterion Eq. \ref{eq:naive}:
 the condition Eq. \ref{eq:Polya} expresses the capability of $K(x,t)$ to adapt and survive the $A(x,t)$ changes while the condition Eq. \ref{eq:naive} merely takes a global, static, average, representative view on the quantities of $A$ and $K$.
  
The property $Pol_{d} = 1$ for  $ d \le 2$ expresses the fact that random walkers $a$ in $d \le 2$ dimensions almost  never travel to infinite distances from their original location. This leads to a rather anomalous behavior of random walkers diffusion in    $ d \le 2$  dimensions which culminates with the fact that in a sense the $a$ random walkers never leave definitively their neighborhood of origin and return eternally again and again to it. This allows the large $k$ 'herds' / 'mountains' / 'islands' that the $a$ agglomerations created to be continuously revisited and reinforced by the $a$'s that originally created them. 
Thus in $ d \le 2$ dimensions, the $k$'s survive even in the most severe conditions: the collective $k$ objects have almost always the chance and time to adapt their position to the very mild and slow changes in the $a$ configuration  that the relation $Pol_{d} = 1$ implies \cite{Louzoun 2007}. 

This is not typical at all: 
in fact for $d > 3$ the $a$'s have a large probability never to return to the region of origin. In fact only about $1/3$ of them do and the rest are lost to 'infinity'.  The condition Eq. \ref{eq:Polya} for the $k$  'herds' / 'islands' / 'mountains' survival becomes non-trivial:
\begin{equation}
s/D_A>0.659563
\label{eq:Polya3}
\end{equation}
Note that while Eq. \ref{eq:Polya3} is a less shocking result than the absolute resilience of the $d<2$ system 
\footnote{It was noted that species choose to live in 2 dimensional environments even if they do have in principle the option of a 3 dimensional space \cite{Vieira 2013}}
, this result is still at complete disagreement with the naive continuum condition Eq. \ref{eq:naive}.  

In conclusion, the system Eqs. \ref{eq:s}, \ref{eq:delta} far from being characterized by the scalar Eq. 
\ref{eq:malthus} presents extraordinary resilience related with the spontaneous emergence of 
growing herds of $k$ around the locations where large enough fluctuations of the number of $a$ above the average $A$ exist. These fluctuations in the $a$'s result in exponentially growing fluctuations in the $k$ herds which often reach macroscopic or at least mezoscopic sizes. In turn these are responsible for the large time fluctuations of $K(t)$.

Thus in order to understand the macro dynamics one has to understand first the nature of the micro  and mezo-granularity represented by the emergent collective objects.

 In the end the possibility of the  $k$  collectives to follow the changes in the growth centers’ location depends on the  $k$  proliferation rate $s$, the $a$’s diffusion rate $D_A$ and the Polya’s constant of the geometry of the space in which the $a$’s and  $k$’s move. Thus Eq. \ref{eq:granularity} expresses the fact that the  $k$’s form collective objects – 'islands' / 'clusters' / 'herds'  – which adaptively identify, follow and exploit to their advantage the fortuitous fluctuations of the $a$ medium. The spontaneous emergence of adaptive, collective objects out of the completely mechanical  $k$’s is what renders the system Eqs. \ref{eq:s}, \ref{eq:delta} anomalously resilient.

The adaptive collective clusters of proliferating individuals $k$ which follow and exploit the gain / growth opportunities offered by the fortuitous temporary agglomerations $a$’s \cite{Louzoun 2003a} have important implications on the characteristic of the time fluctuations of the entire system.

The appearance and disappearance of such collective objects is responsible for large non-self-averaging fluctuations in the system.    For the theorems and for the exact conditions in which these results hold, see \cite{Shnerb 2000}, \cite{Louzoun 2007}, \cite{Challet 2009}. 

Such effects have been described in relation to the rises and falls of globalized economies \cite{Louzoun 2003}, \cite{Goldenberg 2004}, \cite{Huang 2001}, \cite{Challet 2009}.

We will see in the next sections that when interpreted in terms of capital  dynamics, this implies that the different parts of the economy have very heterogeneous intrinsic growth rates and that the system as a whole has very outstanding resilience.

In fact one can say that an economy or ecology does not stand a chance in the absence of heterogeneity. The spontaneous emergence of collective adaptive objects in systems of many autocatalytic 
(proliferating) elements is crucial for their survival. Such collective objects (large wealth accumulations, herds, high-tech integrated labor communities) insure the resilience of the entire system.
They are the carriers and the embodiment of a eventful 'evolutionary' dynamics, never in equilibrium, ever changing in the processes of creation, destruction, growth and decay.

\section{Probability of return to origins, system inertia and fractal intermittent walks}
\label{sec:origin return}

Section \ref{sec:Polyaresilience} studied the properties of random walkers in terms of their probability 
of returning to their original position. This is a powerful method to extract quantitative predictions from 
stochastic models involving random sequences. In addition, it allows the qualitative discrimination between 
various dynamical regimes: 
\begin{itemize}
\item normal (Gaussian 'microscopic' fluctuations), 
\item fractal (Levy \cite{Mandelbrot 1982}, 'scaling fluctuations') and \item intermittent (Zel'dovich \cite{Zeldovich 1987}, catastrophic fluctuations  / extreme events).  
\end{itemize}

In the next subsections we will use those measures in order to relate the granularity of the distribution of $K(x,t)$ for fixed time
to the time fluctuations of its aggregate $K(t)$.
We will find that the size distribution of the large herds characterizing $K(x,t)$'s granularity determines the size distribution of the $K(t)$ fluctuations  \cite{Malcai 1999}, \cite{Solomon 2003}, \cite{Klass 2007}, \cite{Solomon 2001}.

The present section develops the language and tools to explain this connection. 

\subsection{Probability of return to origins, system inertia and fractal / intermittent walks}
\label{sec:Levy Polya}

 The value $Pol_d = 1$ of the Polya constant  is related to  $P_{origin} (2t) = P(x=0,2t)$ the probability for an agent $a$ starting a random walk at origin to be found at origin after $2t$ time steps
\footnote{on a square lattice, the probability of return after an odd number of time steps vanishes $P(x=0,2t+1) =0$.}.
More precisely
\footnote{see  G. Lawler and L. Coyle: Topics in Contemporary Probability and  \url{http://stat.math.uregina.ca/~kozdron/Research/Talks/duke_polya.pdf}}:
\begin{equation}
Pol_d =1 \;  \text{iff} \;  \sum_t P_{origin} (2t) = \infty.
\label{eq:Polyasum}
\end{equation}
For $d=1,2$ this is the case:
\begin{equation}
 P_{origin} (2t) ={(t\pi)}^{-d/2} ; \; \ d=1,2
\label{eq:P0l}
\end{equation}
and thus
\begin{equation}
\sum_t P_{origin} (2t) =\sum_t {(t\pi)}^{-d/2} \rightarrow \infty ; \ d=1,2.
\label{eq:P0sum}
\end{equation}

We learn from Eq. \ref{eq:Polyasum} that the quantity $ P_{origin} (t)$ is very important not just in order to establish the properties of the small fluctuations of a random walk process but mainly for characterizing the extremely large fluctuations - the ones that can bring the system arbitrarily far from its origins. Thus we will see in the Appendinx \ and use in the next Sections that the best way to characterize the large fluctuations of the system is to study $ P_{origin} (t)$:
\begin{itemize}
\item theoretically its time scaling exponent as in Eq. \ref{eq:P0l} connects to the scaling of the sizes of the individual time steps in $K(t)$ and through them to the scaling properties of the clusters, herds  and other collective objects composing the system.
\item practically it allows performing high precision, high reliability measurements which avoid the 
statistics and finite size problems that plague the precision and reliability of the direct measurements of the large fluctuations from the origin \cite{Mantegna 1999}.
\end{itemize}

\subsection{Clusters, $k$-herds and Levy process properties}
\label{sec:Levy herds} 

The quantity $P_{origin}(t)$ introduced in the previous Section  has a crucial role in characterizing the fractal / intermittent / 'cyclic' properties of stochastic systems.

As proven in the Appendix this quantity intermediates between the granular structure of the emergent collective mezoscopic objects in which the microscopic components of the system self-organize and between the time evolution of the system as a whole.
More precisely consider a generic system where: 

\begin{enumerate} 
\item the system size $K(t)$ is the sum of the sizes $K_i (t)$ of the $i$ collective objects:
\begin{equation}
K(t)= \sum_i K_i (t).
\label{eq:Kisum}
\end{equation}
Consequently the unit steps $\Delta K(t)$ in the time evolution of $K(t)$ are the unit steps $\Delta K_i$ of the individual $K_i$'s (assuming the dynamics is asynchronous i.e. the $K_i$'s are not updated at exactly the same time / 'synchronous').
\item As a result of the microscopic autocatalytic dynamics
(e.g. Eq. \ref{eq:s}, \ref{eq:delta})  the sizes of the components when ordered (in decreasing order) according to their rank (largest $K_i$ first) respect a Pareto-Zipf power law:
\begin{equation}
K_i \sim i^{-\alpha}.
\label{eq:KiZipf}
\end{equation}
\item Due to the same autocatalytic character of the microscopic dynamics, the short time changes $\Delta K_i$ of the $K_i$'s  are distributed by the same Pareto-Zipf power law.  This  can be written equivalently to 
Eq. \ref{eq:KiZipf} in terms of the probability of a step $\Delta K_i$ to be larger than a certain $StepSize$ value:
\begin{equation}
Prob_{Pareto} (\Delta K_i > StepSize) \sim (StepSize)^{-\alpha}
\label{eq:KiPareto}
\end{equation}
Taking into account  Eq. \ref{eq:Kisum}, this means that the time evolution of $K(t)$ is a Levy flights process;
i.e. a random walk with steps distributed by a Pareto power law:
\begin{equation}
Prob_{Pareto} (\Delta K > StepSize) \sim (StepSize)^{-\alpha}
\label{eq:KPareto}
\end{equation}
\end{enumerate}

In these conditions (1-3), according to the Levy and Mandelbrot analysis \cite{Mandelbrot 1982} (see Appendix for the details of the argument), the time variation of $K(t)$ over a time interval of $t$ steps is a Levy distribution $L_{\alpha} (\Delta K, t)$ with a width that expands in time as:
\begin{equation}
\sigma_{Levy} (\Delta K, t) \sim t^{1/{\alpha}}; \ for \ \alpha < 2 
\label{eq:sigmaLevy}
\end{equation}
which (as explained above) is reflected in the scaling in time of its central peak as:
\begin{equation}
P_{Levy} (\Delta K(t)=0,t) = P_{origin \ Levy} (t) \sim 1/\sigma_{Levy} (\Delta K, t) \sim  t^{-1 / \alpha}; \ for \ \alpha < 2
\label{eq:PoriginLevy}
\end{equation}

 Thus the behavior of the $K(t)$ Levy flights process depends on the actual values of $\alpha$ i.e. of the size of the elementary objects (large lumps of individual capital, large firms, investor herds, production sectors, highly productive geographical locations) that determine the individual steps size distribution
\footnote{Note however that above a certain amplitude, the very large fluctuations are disturbed by finite size effects and have a very different exponent of the power law \cite{Mantegna 1999}.}.

\begin{itemize}
\item For {\bf $\alpha > 2$ } one recovers the Gaussian result.
\item For $1< \alpha <2$ the fluctuations are fractal (Levy stable) and reach larger scales 
(but still limited by the finite size of the largest individual objects).
The probability for the system to be at the original state after a time $t$ decreases much faster then in
the Gaussian case. More precisely the speed by which the system gets away from the original value is dictated  by the same exponent $\alpha$ that characterizes the step size distribution Eq. \ref{eq:KiZipf} (Eq. \ref{eq:PstepLevy} in the Appendix) and is measured by the decay speed of the central peak Eq. \ref{eq:PoriginLevy} (Eq. \ref{eq:Levy width} in the Appendix).
\\ {\bf Fig. 1} demonstrates the remarkable agreement between the theoretical prediction and the empirical data by comparing the exponent in Eq. \ref{eq:KiZipf} (size of the wealth of the largest investors) vs. the exponent in Eq. \ref{eq:PoriginLevy} (the time dependence of the central peak of the market index fluctuations).
\item For $\alpha <1$ the fluctuations are so strong that essentially the system is completely dominated by the largest steps sizes and the dynamics is dominated by intermittent singular events (see Sections \ref{sec:Predictions} and \ref{sec:empirical}) similar with the economic cycles of crises and booms.
\end{itemize}

It is very encouraging that the same theoretical formalism, by just changing the value of a parameter $\alpha$ is capable to fit, explain and predict such a wide range of empirical phenomena, from microscopic short time fluctuations, via fractal fluctuations spanning a wide range of time and amplitude scales, to very rare extreme fluctuations with macroscopic systemic dimensions both in amplitude and in time.

\begin{figure}
  \includegraphics[scale=1.00]{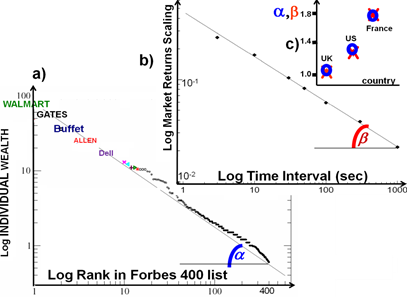}
\caption{ {\bf  Granular structure of individual wealth / capital distribution determines the size distribution of the financial system time fluctuations} 
\\ This is an economics realization of the wider result that in microscopic Eqs. \ref{eq:s} - \ref{eq:delta} 
and mezoscopic  \ref{eq:Kisum}-\ref{eq:KiZipf} or \ref{eq:Kcomponents}-\ref{eq:AKmatrix}
discrete stochastic models: the granularity of the $k$ clusters / herds / collective objects (sub-figure a) determines the time fluctuations of total quantity of $k$' in the entire system $K(t)$ (sub-figure b).\\
\\ {\bf Figure 1a)} represents the granular structure of the wealth distribution in terms of the individual wealth of the wealthiest people in US (Forbes 400). It is found that it fulfills a Pareto-Zipf power law distribution 
$W(N) \sim N^{-1/\alpha}$ where $N$ is the rank of the individual in the Forbes list and $W(N)$ is their wealth.
\\ {\bf Figure 1b)} represents the fractal scaling properties of the returns $\Delta K(t)$ distribution $P(\Delta K,t)$ in terms of the probability $P_{origin} (t) =P(\Delta K(t)=0,t)$ of finding the current market index $K(t)$ at the same value after a time interval $t$. One finds that $P_{origin} (t) \sim t^{-1/\beta}$ which implies cf. 
\ref{eq:PoriginLevy}: $\sigma (t) \sim t^{1/\beta}$ (note that this does not imply necessarily that the very long tail of the distribution decays by the same exponent \cite{Mantegna 1999}).
\\ {\bf Figure 1c)} compares the empirical values of the 2 exponents in 3 different countries:
\\ - blue circles represent $\alpha$ (exponent of the individual agents' capital distribution) 
\\ - red crosses represent  $\beta$ (exponent of the fluctuations of the market index).
 One finds that $\alpha = \beta$ thereby validating empirically the predictions  cf. Eqs.  \ref{eq:KiZipf}, \ref{eq:PoriginLevy} (\ref{eq:Zipf law}, \ref{eq:Levy sum} in the Appendix).
}
\label{fig:Pareto}      
\end{figure}

In the present paper we will give an interpretation of the results of \cite{Shnerb 2000}, \cite{Louzoun 2007} in a form closer to Aoki’s work and keep only the autocatalytic features necessary for the creation-destruction process. Moreover, rather than dealing directly with the 'microscopic agents' based model, we take a coarser - \emph{'mezoscopic'} - view, which is an intermediate level description of the Eqs. \ref{eq:s}, \ref{eq:delta} system, between:
\begin{itemize}
\item the original microscopic one (studied in \cite{Shnerb 2000}, \cite{Kesten 2002}) consisting of a macroscopic number of individual microscopic agents and 
\item the classical macroscopic one Eqs. \ref{eq:malthus}, \ref{eq:Malthus solution} \cite{Malthus 1798} in which the entire system is represented by  a single macroscopic variable $K(t)$.
\end{itemize}
Rather than not aggregating at all or aggregating to a single global representative agent, we aggregate to the level of the emergent herds found in the microscopic model.

The hypothesis described in the following sections that the complex economic cycles dynamics resulting from the myriads of microscopic interactions can be reduced to discontinuous events that correspond to the appearance / growth and disappearance / shrinking of the dominant herd / economic sector is very non-trivial and may be over-simplifying 
\footnote{Popper called science "the art of over-simplification".}. 

However it is strongly supported both by the analysis of the first principles system Eqs. \ref{eq:s} , \ref{eq:delta} and by the confrontation with the data (see Section \ref{sec:empirical}).

Moreover it gives formal quantitative support to deep ideas formulated previously only in words by Schumpeter and Minsky. 
To develop this effective mezoscopic formulation in an growth economics context, we start below from the simplest  endogenous economy growth model \cite{Romer 1986}.

\section{The AK model and its multi-component heterogenous extensions}
\label{sec:Logistic AK}
The $AK$ model describes the growth process starting from two assumptions: 
\begin{enumerate}
\item that the current flow of output goods $Y$ is proportional to the stock of capital $K$ (which aggregates physical, human and intellectual capital):  
\begin{equation}
Y=AK
\label{eq:AK}
\end{equation}
where $A$ is a constant \cite{Mankiw 2006}, and
\item that the capital accumulation in time $dK/dt$ as a difference between the investment (assumed equal to the savings which in turn are assumed proportional to $Y$) and the capital depreciation (assumed to be proportional to $K$):
\begin{equation}
\frac{dK}{dt}=sY-\delta K.
\label{eq:dKdt}
\end{equation}
\end{enumerate}
By substituting $Y$ from Eq. \ref{eq:AK} into Eq. \ref{eq:dKdt}, one obtains the differential equation describing the growth of the economy in the $AK$ model: 
\begin{equation}
\frac{dK}{dt}=(sA-\delta) K=gK.
\label{eq:Eq25}
\end{equation}
where $g$ is defined as in Eq. \ref{eq:growth rate} with of course new, different interpretations of the $K$ variable and the  $s$, $A$, $\delta$ constants.

This establishes the formal equivalence of the Malthus Eq. \ref{eq:malthus} to the endogenous growth model Eq. \ref{eq:Eq25}. In fact, as mentioned above, we chose intentionally 
notations that while non-standard in the agent-based models, are familiar to the $AK$ practitioners.

This is a mathematical equivalence between the equations expressing life re-production and the equations expressing capital reproduction. They both share the autocaliticity property that the variation $dK/dt$ of $K$  is proportional to $K$ itself. We will argue below that this autocatalytic behavior is the very basis of the emergence of most macroeconomic phenomena out of microeconomic individual interactions. 

As with Eqs. \ref{eq:malthus}, \ref{eq:Malthus solution}, the simplest endogenous growth model Eq. \ref{eq:Eq25} where $A$ and $K$ are scalars cannot describe systems which consist of heterogeneous components: very often a macroeconomic system consists of many interacting subsystems with different natural resources, different labour prerequisites and different socio-economic structures. The inhomogeneity  can be among regions in real geographical space \cite{Yaari 2008} or among sectors in the space of products \cite{Hidalgo 2007} or among the economic sectors \cite{Leontief 1947}. Thus rather than lumping all capital in one number, $K$ has to be generalized to a collection of interacting heterogenous parts \cite{Aoki 1999}, each with a growth rate governed by different coefficients and interacting among themselves.
In fact we will find in this paper that the granularity of the emergent collective objects that compose the system determines the characteristics of the time evolution of the system as a whole: the individuals wealth distribution determines the fluctuations of the stock market indices Fig. 1, the economic sectors determine the GDP cycles behavior Fig. 5, the geographical regions determine the system behavior after shocks Fig. 6 etc.

 The interaction between the microeconomic heterogenous elements, leads to a wide range of non-trivial consequences: self-organization, emergence of adaptive collective objects and anomalous resilience \cite{Shnerb 2000}.

The choice of the nature and scale of the parts composing $K(t)$ and of the space in which they act is a crucial modeling decision which has to be adapted to the specific application.
 In \cite{Shnerb 2000} the discrete agents $k$ moved in a space background populated by growth agents $a$. The agents $k$ and $a$ as well as the space in which they diffuse admit an extensive range of interpretations and of coarser , mezoscopic representations. 
In this paper we will refer only to three examples:
\begin{itemize}
\item the space is the network of individual investors, the number of $k$ on a node is the investor's capital, the links represent capital flow between the investors and the number of $a$'s on a site is proportional to the returns the investor is getting ($k$ proliferation). This  model is the micro-economic 
basis for the mezo-economic effective model discussed in  Section \ref{sec:Levy walk}.  
\item the space is the network of geographical locations, the $a$'s are people's capabilities, $k$ are companies. This is the microeconomic foundation for the Poland post-liberalization mezo-economic discussion in Sections \ref{sec:Predictions}-\ref{sec:empirical}.
\item the space is the supplier-client production network. Links represent client-supplier relationships. The  amount of $k$'s on a given node  is  the capital of the company and the $a$ its productivity. 
The mezo-economic counterpart is the space where nodes $i$ are economic sectors, $K_i$ are their capital, links are capital transfer lines $A_{ij}$'s. $A_{ii}$ are intrinsic growth rates of the sectors $i$. This is the basis for the discussion in \cite{Dover 2009} \cite{Challet 2009}, Section \ref{sec:Predictions} and Figs. \ref{fig:double} and \ref{fig:Balkan}.
 \end{itemize}
 
As mentioned above, the analytic treatment of systems with infinite ('thermodynamic limit') or very large number of components requires the use of heavy combinatorial, renormalization group or branching random walks techniques. This often makes less evident the crux of the matter.

The spontaneous emergence of the mezoscopic adaptive self-organizing objects is the key to the outstanding resilience of the economic systems. Political regimes that try to eliminate the heterogeneity 
(inequality) in the system and / or the fluctuating nature of its time evolution, prepare the way to their own demise.

 In the remaining sections we introduce a mezoscopic representation of the systems where the herds / clusters / individual capital lumps are considered as the elementary objects. Rather than following up their spontaneous emergence from the microscopic interactions as in Sections 4-5 we include them explicitly in the model from the beginning and study their consequences.

Thus, instead of splitting the total macroscopic $K$ into a multitude of heterogeneous interacting agents $k$, we will only generalize $K$  to a vector   $\vec{K}$ whose components $k_i$ represent the capital in different parts of the economy:
 \begin{equation}
\vec K (t) \equiv (k_1, k_2, ..., k_n) .
\label{eq:Kcomponents}
\end{equation}
Appropriately, as opposed to Eqs. \ref{eq:malthus}, \ref{eq:Eq25},  $\vec{G}$  is not a scalar anymore, but it is a matrix of elements that represent inflow and outflow between the elements of the vector $\vec{K}$.
 Thus we reach as a hetrogenous generalization of Eq. \ref{eq:Eq25} the system of linear equations:
\begin{equation}
\frac{d \vec{K}}{dt}=\vec{G} \cdot \vec{K}(t).
\label{eq:AKmatrix}
\end{equation}

In the case in which $\vec{G}$ does not depend on time, the formal solution is: 
\begin{equation}
\vec{K}(t)=e^{\vec{G} t} \vec{K}(0).
\label{eq:MatrixSolution}
\end{equation}
Or more explicitly,
\begin{equation}
\vec{K}(t)=\sum_{i=1}^{N}  <\vec{K}(0), \vec{u}_{i} >  e^{\lambda_{i} t} \vec{u}_{i},
\label{eq:Solution}
\end{equation}
where $N$ is the dimension of $\vec{G}$, and  $\lambda_{i}$ and $\vec{u}_i$ are its eigenvalues and eigenvectors. However, we will also discuss here  the interesting cases where $\vec{G}$ changes in time.

The components of $\vec{K}$ can represent economic clusters, sectors, geographic regions, investors or companies. Thus the elements of the $\vec{G}$  matrix can represent capital (wealth) transfer between either geographical domains or transfers between economic sectors, or the gains of various companies or investors. In the case where the components are different economic sectors, the matrix $\vec{G}$  is similar to the Leontief matrix (except that it includes endogenously the demand term by including labor, consumption etc).  In this interpretation, each column of  $\vec{G}$ reports the monetary value of a sector’s (or geographical region) inputs and each row represents the value of its outputs. The  $\vec{K}(t)$ components can still be connected to the (aggregated) capital in each component of the economy but the connection to the GDP is more complicated because it depends on the details of the payments within and between the various components. We will thus consider the aggregated capital $K_{tot}(t)$ to be the sum over the $\vec{K}$ components:
\begin{equation}
K_{tot}(t)=\sum_{i=1}^N k_i(t)
\label{eq:Ktot}
\end{equation}
where $k_{1}(t)$, $k_{2}(t)$, $\cdots$, $k_{n}(t)$ are the components of the vector $\vec{K}(t)$. 

Assuming similarity to Eq. \ref{eq:AK}, the GDP $Y_{tot}(t)$ is still proportional to $K_{tot}(t)$ but in a rather complicated way that is largely missed by the coarse representation of the payments that  $\vec{G}$ contains.
This similarity extends to the {\ it changes} in the individual capitals $k_i (t)$:   the change in the total wealth $\Delta K_{tot}$ is the sum of the (asynchronous) changes in the individual wealths $\Delta k_i$ which in turn are proportional (in a complicated matriceal, stochastic way) to the $k_i$ themselves.

\section{ Granular individual wealth,  intermittent macroeconomic returns, Pareto law and Levy walks  }
\label{sec:Levy walk}

As mentioned above, in the generic case the matrix $\vec{G}$  varies in time. The fact that the microscopic agents $a$ are dynamic reflects into the fact that $\vec{G}$  is not a constant either. In particular, because of the exponential in Eq. \ref{eq:MatrixSolution}, any additive change in  $\vec{G}$ leads to a multiplicative change in $\vec{K}(t)$. Thus the changes in the  individual wealths  are proportional to the wealths themselves via a random factor. This has been confirmed in \cite{Klass 2006}. 

The last paragraph implies that the changes in the total wealth constitute a sequence of jumps proportional to the wealths of the individual economic players.  
This prediction has been formalized in \cite{Malcai 1999}, \cite{Solomon 2003} in a relation between the granularity of the individual wealths at a certain time and the market index fluctuations in the same period.
In terms of the present paper this means that the exponent in Eq. \ref{eq:KiZipf} (Pareto-Zipf distribution of the wealth of the largest investors) and the exponent in Eq. \ref{eq:PoriginLevy}
(the exponent characterizing the  fractal distribution of returns of the stock market index) are equal. This has been very precisely confirmed by subsequent empirical measurements (Fig. \ref{fig:Pareto}).

Beyond the academic interest of having empirical confirmations to theoretical models, this relation between the measure of inequality in the wealth (Pareto exponent) and the measure of instability in the financial markets (fractal exponent of the market fluctuations) has important practical and social interest. It shows that social inequalities are not only morally and socially problematic, but they are endangering the interests of the capital itself. Limiting the wealth inequality is a capital interest not less that an interest of the humans involved in the economy.

If one interprets the individual capital clusters $K(x,t)$  as the wealth of the individuals and the stock market index as a measure of the capital $K(t)$ in the system, then one predicts that the time changes in $K(t)$ are proportional to the relative sizes of the individuals’ wealth. 
This prediction is validated by the subfigures (a), (b) and (c) in Fig. \ref{fig:Pareto}:
\begin{itemize}
\item[(a)]	represents the distribution of wealth sizes of the (richest) individuals in US. It turns out this is close to a power law with Pareto exponent $\alpha$ (Eq. \ref{eq:KiZipf}). 
\item[(b)]	represents the height ($\sim$1/width) of the distribution of returns in the (S\&P) stock market index as a function of the time interval on which the returns are measured. It is well approximated by a power law with fractal exponent $\beta$ (Eq.  \ref{eq:PoriginLevy}).
\item[(c)]	represents the values of $\alpha$ (blue circles) and $\beta$ (red crosses) for 3 western economies. As predicted by the theory, the 2 exponents, while very different for the different countries, are equal for each country, $\alpha=\beta$. Thus the granularity of wealth at a given time determines the amplitudes of the fluctuations of the total wealth of the system.
\end{itemize}
In terms of the possible values for $\alpha$ one distinguishes 3 regimes:
\begin{enumerate}
\item For  $\alpha > 2$  one recovers the Gaussian random walk.
The granularity of the individual wealths is not felt at the macroscopic level because there are no 
individuals with macroscopically significant wealth / capital.
Thus, the fluctuations of the total wealth  in the system (e.g. market index) are the result of many microscopic contributions which cancel one another in great measure and keep the system free of macroscopic fluctuations. The system is very stable as a result of a rather egalitarian wealth distribution among the players / traders / investors.
\item for $1< \alpha <2$ the fluctuations 
are larger and the probability to be at origin after a time $t$ decreases much faster then in
the Gaussian case. More precisely the speed by which the system gets away from the original value  Eq.  \ref{eq:PoriginLevy} (in the Appendix \ref{eq:Levy sum}) 
is dictated by the same exponent $\alpha$ that characterizes the wealth size distribution given by Eq. \ref{eq:KiZipf}) (in the Appendix \ref{eq:Zipf law}). In particular the fluctuations have a divergent second momentum (square standard deviation). Thus, while the largest individual players have substantially smaller wealth than the entire wealth in the system, their anomalous power law distribution of wealth influences the market fluctuations making the  volatility of the order of the macroscopic system wealth rather then the naively expected $1/\sqrt{number  \ of \ individual \ agents}$. 
\item for $\alpha <1$ the fluctuations are so strong that essentially the system is completely dominated by the largest steps sizes. For this case virtually the entire wealth of the system is concentrated in a very small number of players and their individual actions directly impact on the market index. Thus large wealth inequality ends up in very dangerous financial market instability \cite{Huang 2001} \cite{Louzoun 2003}.
\end{enumerate}
While most of the economies are in the case 2 when $1< \alpha <2$ (as seen in Fig. \ref{fig:Pareto})
there are examples of countries like Argentina, Brasil, former USSR countries after liberalization in which at least temporarily  the inequality reached  $\alpha <1$  and resulted in financial mayhem \cite{Huang 2001} \cite{Louzoun 2003}

Moreover, as seen in the next sections, the $\alpha <1$ regime in which the most of the  system is 
dominated by a very small number of the largest components is capable to describe and predict the 
typical behavior of economic cycles .

 In fact the present treatment in which different regimes differ 
by just the value of a single parameter ($\alpha$) blurs - and even gives the hope of erasing - the
boundaries between micro- and macroeconomics. If successful, this would be truly a realization of Aoki's vision.

\section{Schumpeter's creative destruction and the universal 'crossing exponentials' pattern}
\label{sec:crossing}

One can go now one step beyond the statistical connection between the macroeconomic dynamics and the economy granular structure:  one can connect the time variations in $\vec{K}$  to specific events in $\vec{G}$ which affect deferentially the $\vec{K}$ components.

In \cite{Louzoun 2007} it was proven analytically that the microscopic stochastic $a$ density fluctuations lead to the creation and destruction of macroscopic collective objects. 

The growth / shrinking of those objects turned out to be responsible for a very specific universal pattern 
that appears ubiquitously in the time evolution of the global system. We call this pattern the 'crossing exponentials'. The 'crossing exponentials' pattern is the quantitative and visual expression of the Schumpeter 'creative destruction' idea. 

In the following we will express mezoscopically this empirically observed  economic dynamics  as the effect of sudden discontinuous changes in the matrix  $\vec{G}$.  These jumps in $\vec{G}$ may correspond to the changes produced by innovation or any other endogenous or exogenous changes in the economy.  Obviously, such changes affect differently the different components of the economy and even initiate dramatic events like the dismiss or the inception of entire economic sectors or production technologies. 

The fact that these changes  can be reduced at the macro-economic scale to \emph{discontinuous changes in a low-dimensional matrix $\vec{G}$} is a non-trivial hypothesis (suggested by the above mentioned previous microscopic studies and by Schumpeter's analysis). This hypothesis has to be confronted with the empirical data in the same way in which the hypotheses underlying the microscopic model Eqs. \ref{eq:s}, \ref{eq:delta} have been confronted with the empirical growth data in \cite{Yaari 2008}. 

To do this, we expose in the next sections the implications of the mezoscopic $\vec{G}$ model of Schumpeterian  creative destruction and compare them with the empirical data. 
The main predictions, confirmed by the empirical data will be: 
\begin{itemize}
\item	Between the shocks of the matrix $\vec{G}$  the  model Eq. \ref{eq:AKmatrix} represents a quasi-stationary economy: if the matrix  $\vec{G}$ is constant for a long enough time, the economy reaches a steady state in which it grows at a constant rate, preserving the ratios between its sectors sizes.
\item	 Upon a generic change in the matrix $\vec{G}$  the system enters an exponentially decaying phase of destruction with various components having different different growth / decay rates. 
\item Eventually the system reaches a new steady state of uniform growth rate for the entire economy (still with large differences in the absolute GDP per capita of various components). 
\end{itemize}
We present below the simple case of a 2x2 $\vec{G}$  matrix.
While in detailed empirical studies we considered matrices as large as 3000x3000 (e.g. for the Polish post-liberalization application Fig. \ref{fig:Poland}) or spatially extended systems with millions of agents $a$ and $k$ cf. Eqs. \ref{eq:s}, \ref{eq:delta}, as in \cite{Louzoun 2003}, \cite{Louzoun 2003a}, the 2x2 example provides a surprisingly good fit (Figs. \ref{fig:crossing}, \ref{fig:double}) to many real economic events (e.g. Fig. \ref{fig:Balkan}).Remarkably, it does it very economically in the Occam razor sense by invoking a very small number of parameters that can be readily calibrated \cite{Challet 2009}.  In the cases where we were able to get enough detailed data the explanation for this remarkable simplicity is the dominance of the exponentials over other functional dependencies in the system. More precisely, it has been found in \cite{Dover 2009} that after a shock, the economic sectors can be lumped to a good approximation in 2 parts / herds:
\begin{itemize}
\item the part that dominated the economy until the shock and undergoes destruction following the shock.
\item a part that was rather obscure before the shock but it is the main beneficiary of the changes and undergoes Schumpeter creative growth in the aftermath.  
\end{itemize}

If the 2 parts are not significantly disjoint, the shock is mild or even not deserving the classification as a shock.

If the 2 parts are disjoint to a significant degree, then the dynamics of the system is dominated by 2 processes taking place in parallel:
\begin{itemize}
\item The part of the economy that is favored after the shock starts growing exponentially. By definition, having a limited overlap with the old dominant part, it starts with a size that is a small part of the entire economy. However given its growth with the new, dominant growth rate, it eventually becomes the largest part of the economy, dictating in particular the overall growth rate of the entire economy.  
\item The part of the economy that dominated the economy before the shock undergoes an exponential decay. While at the beginning it constituted the largest part of the economy and imposed the overall growth / decay rate, at some stage it shrinks to a sub-dominant position. Its growth may be eventually restored only in as far as it interacts with (and benefits from transfers / business from) the new growing, dominant part of the economy.
\end{itemize}
This trivial example gives an additional illustration of the following facts.
\begin{itemize}
\item If $\vec{G}$  is a matrix, even if the average of its elements is a negative constant, one can get rather than an exponential decay in $K_{tot}(t)$ as suggested by Eq. \ref{eq:growth rate}, an overall exponentially growing trend for $K_{tot}(t)$ in Eq. \ref{eq:convexity} and Eq. \ref{eq:Solution}. 
\item Immediately after changes in $\vec{G}$ (representing  punctuated interesting events, such as spectacular crises and reforms) one observes a decreasing exponential decay even if on long range the growth is accelerated. 
\item  As opposed to the typical depiction of the economic cycle fluctuations where both the maxima and the minima are smooth, the model predicts that the {\bf maxima are cusps}.  In between them, $K(t)$ is a sum of 2 exponentials (corresponding to the old and new dominant economy components e.g. Fig. \ref{fig:Balkan}) with a smooth bottom.
\end{itemize}

We will call this generic scenario the \emph{Schumpeter 'crossing exponentials' after-shock effect}. 
Even in the cases where one considers more than 2 parts, the corrections are negligible because the largest exponentials before and after the shock will dominate \cite{Yaari 2008}.
 Even in the cases where the shock advantages the new sector / region without suppressing the first, one still has a crossing  in the relative weight of the old and new sectors and an eventual dominance of the growth rate of the new sector. Such shocks are less dramatic than the usual ones in which the new sector is advantaged on the expense of the old sector. However, with enough (quarterly)  time resolution one can detect in the empirical data  that the growth of economies is generically a sequence of such crossings \cite{Challet 2009}.   

It is still non-trivial that the transition between regimes can be described by punctuated changes in the matrix $\vec{G}$ rather than a smooth evolution of it. We connect it with the existence, emergence and collapse of the adaptive collective objects discovered in the context of the microscopic models \cite{Shnerb 2000}, \cite{Louzoun 2003}.

The 2x2 model described in the next section translates the intuitive but brilliant narrative of Schumpeter into a mathematically precise formulation which takes into account in a quantitative way the overlap between the new and old sectors favored by the dynamics before and respectively after the shock. As it turns out, the predictions of this simplified model are quantitatively confirmed by the empirical data. 

An effect which is missed (or rather included by hand / by dictum) through the reduction of the model to just 2 components is the extreme resilience of the system with many components. Indeed as found in \cite{Shnerb 2000} and \cite{Kesten 2002}, when the number of components diverges to infinity, the system’s resilience diverges too (at least in the 2D geographical case).  

This is because the probability $P$ that with a random choice of the new $\vec{A}$ and $\vec{G}$  matrices one still has at least one growing component approaches $P=1$. This implies that even economic systems with zero-intelligence investors survive if they are diverse enough and if the capital is reinvested in the neighborhood of the sector in which it was created.

 In the 2x2 example, this effect has to be included 'by hand': in order to obtain growth, at least one of the 2 sectors (the new one) has to be assigned a positive intrinsic growth rate. Thus, unlike the many-agent model Eqs. \ref{eq:s}, \ref{eq:delta} where the emergence of a positive growth subsystem
$S$  is guaranteed cf. Eq. \ref{eq:growth g} with probability of 1, Eq. \cite{Shnerb 2000}, in the 2x2 $\vec{G}$ matrix model of creative destruction  this has to be built in as one of the assumptions.

Still the agreement of its predictions with the empirical measurements makes this model useful both conceptually and as a regulation, forecast and policy tool. In particular the dependence of the dynamics on the off-diagonal terms allows to find the optimal balance between attenuating the pains of the short range recession (by supporting the old part of the economy) without sacrificing the long term recovery (which depends on allowing the new part to take off as soon as possible). This is explained below and in Fig. \ref{fig:double}.

\section{Quantitative predictions of the Schumpeter creative destruction model}
\label{sec:Predictions}

\begin{figure}
  \includegraphics[scale=0.5]{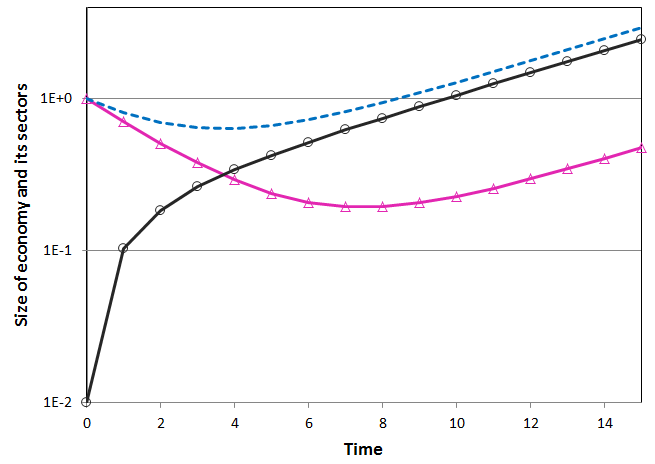}
\caption{{\bf Aftershock time evolution of the 2x2 Schumpeter creative destruction model}
\\ The graphs describe the development of an economy with two sectors after an initial shock. 
 The intrinsic growth rate of the new sector is positive $g_{22}=0.15$ while the one of the old sector is negative  $g_{11}=-0.35$. Their average growth rate is is negative $g =-0.1$ which means that a scalar model aggregating them would predict the decay of the economy. The transfer terms between sectors are  $g_{12}=0.1, g_{21}=0.1$. 
The initial conditions are described by the vector  $\vec{K}(0)  = [k_1 (0)=1, k_2(0)=0.01]$. We take the initial ratio of new  to the old component  0.01 because as it turned out in the Poland application (Fig. \ref{fig:Poland})  the number of counties leading the 'new' economy was 16  while the ones representing the 'old' decaying economy was about 100 larger.
\\ The graphs display the main universal characteristics of the 2x2 Schumpeter creative destruction model:
\\ A. Immediately after the shock, the growth rates diverge: the old sector $k_1(t)$ (pink triangles) decreases exponentially (rougly with decay rate $g_{11}$) while the new sector $k_2(t)$  (black circles) increases exponentially (rougly with decay rate $g_{22}$).
\\ B. The 2 exponentials cross at the bottom of the crisis (the minimum of $K_{tot}(t)= k_1(t) +k_2(t)$
represented by the blue interrupted line).
\\ C. After some time, the growth rates of the 2 sectors $dk_1(t)/dt$ and $dk_2(t)/dt$  allign and both equal $\lambda_{max}$ (the largest eigenvalue of the matrix $\vec G$. 
\\ D. However, the difference between the values of $k_1(t)$ and $k_2(t)$  increases exponentially keeping their ratio roughly at the level of the capital transfer term $ k_1(t) / k_2(t) \rightarrow g_{12}$  from the new sector to the old one.
 }
\label{fig:crossing}       
\end{figure}

\begin{figure}
  \includegraphics[scale=0.5]{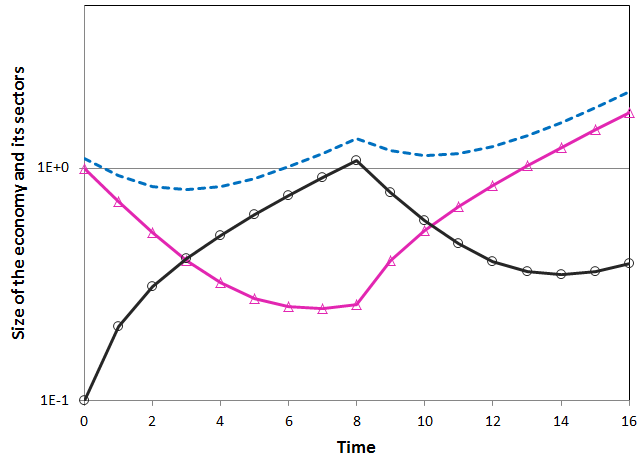}
\caption{ {\bf Shock in the middle of the recovery}
\\The hypothesis that underlies the matrix Schumpeter creative destruction model is that the microscopic agents model Eqs. \ref{eq:s}, \ref{eq:delta} dynamics is dominated by the appearance and disappearance of the dominant herds of $k$. In the mezo-economic level, where the microscopic agents are aggregated in herds rather than in a global representative agent this is represented as a change in the growth matrix $\vec {G}$. The present graph starts with  $K(0)=[k_1(0)=1, k_2(0)=0.1]$, and  $\vec{G}= [g_{11}=-0.35, g_{12}=0.1, g_{21}=0.1, g_{22}=0.15]$ and develops the usual cross-exponential pattern. However, at $t=8$, one modifies the growth matrix to $\vec{G}=[g_{11}=0.15, g_{12}= 0.1, g_{21}= 0.1, g_{22}=-0.35]$. This sharply stops the current cross exponential pattern and starts a new one.
\\ The signal is very sharp and unambiguous: as opposed to the typical depiction of the economic cycle fluctuations where both the maxima and the minima are smooth, the 2x2 Schumpeter creative destruction model Eq. \ref{eq:AKmatrix} predicts that the {\bf maxima are cusps}.  In between, $K(t)$ is a sum of 2 exponentials (corresponding to the old and new dominant economy components) with a smooth bottom.
\\ As seen in Fig. \ref{fig:Balkan} , the empirical data greatly validate this.
\\ This pattern is also quite universal: it has been found  also in real estate, population dynamics and many other systems.
 }
\label{fig:double}       
\end{figure}

One of the difficulties in following the ideas underlying the models that bridge between the micro and macro levels is the rather involved mathematical formalism. In this section we will try to avoid it. We will explain some of the effects through the intermediary of an effective model that lumps the agents in just two sectors.   In spite of this gross oversimplification the similarity of its predictions (Figs. \ref{fig:crossing} and \ref{fig:double}) to the data (Fig. \ref{fig:Balkan}) is still remarkable. 

Such a reduction of the number of parameters is often suggested by the data: in the original analysis of the Polish economy after liberalization,  the behavior of the 3000 counties was grouped in 6 herds according to their education level. Of course a priori one does not know the necessary and sufficient level of resolution for a particular application. A posteriori one could represent the post-liberalization dynamics in terms of just two groups / herds: 
\begin{itemize}
\item the 'new' sector flourishing after liberalization (16 counties with 11.5 or more schooling years per capita), and 
\item the 'old' sector hit by liberalization (most of the other counties).
\end{itemize} 
The economy of a country with just two sectors is represented cf. Eq. \ref{eq:AKmatrix} by a differential 2x2 matrix equation that can be decomposed in its orthonormal components:
\begin{equation}
\frac{d \vec{K}(t)}{dt}=< \vec{K}(t), \vec{u}_1 > \lambda_1 \vec{u}_1 + < \vec{K}(t), \vec{u}_2 > \lambda_2 \vec{u}_2
\label{eq:2 sec solution}
\end{equation}

where $\vec{u}_1$, $\vec{u}_2$ are the 2 eigenvectors of the 2x2 matrix $\vec{G}$. $\lambda_1$ and $\lambda_2$ are their respective eigenvalues. To obtain Eq. \ref{eq:2 sec solution} from Eq. \ref{eq:Solution}, we have used: 
\begin{itemize}
\item the fact that one can decompose the vector $\vec{K}$ in terms of the orthonormal basis $\vec{u}_1$, $\vec{u}_2$ $(\vec{u}_1 \perp \vec{u}_2, <\vec{u}_1, \vec{u}_2> = 0, <\vec{u}_1, \vec{u}_1> = 1, <\vec{u}_2, \vec{u}_2> = 1)$,
\begin{equation}
\vec{K}(t)=<\vec{K}(t), \vec{u}_1> \vec{u}_1+ <\vec{K}(t), \vec{u}_2> \vec{u}_2
\label{eq:orth}
\end{equation}
\item The fact that $\vec{u}_1$  and $\vec{u}_2$  are eigenvectors of  $\vec{G}$:
\begin{equation}
\vec{G} \cdot \vec{u}_{1/2} = \lambda_{1/2} \cdot \vec{u}_{1/2}
\label{eq:eigenvectors}
\end{equation}
\end{itemize}
The analytic solution of this linear homogenous 2x2 differential system Eq. \ref{eq:2 sec solution} is plotted in Fig. \ref{fig:crossing}  : 
\begin{equation}
\vec{K}(t)=<\vec{K}(0), \vec{u}_1> e^{\lambda_1 t} \vec{u}_1+ <\vec{K}(0), \vec{u}_2> e^{\lambda_2 t} \vec{u}_2,
\label{eq:analytical}
\end{equation}
where the $\vec{K}(0)$ vector represents the initial conditions before the shock. Assuming that before the shock the 'old' sector $k_1(t)$  has been dominant, and the 'new' sector $k_2(t)$  has been undeveloped implies $k_1(0)>>k_2(0)$.

For definiteness, we also assume that the largest eigenvalue of  $\vec{G}$ is  $\lambda_{max}=\lambda_2 > \lambda_1$. This is natural because by the choice $g_{11}<0$, $g_{22}>0$  the eigenvector  $\vec{u}_{max}=\vec{u}_2$  is closer to the vector representing the 'new' growing sector direction $(0, 1)$ than to the 'old' $(1,0)$.

Let us assume for clarity and in accordance with the discussion above that the component of $\vec{K}(t)$  representing the 'old' sector $k_1(t)$   has after the shock a negative
 intrinsic growth rate  $g_{11}<0$ while the 'new' sector $k_2(t)$  has a positive intrinsic growth rate $g_{22}>0$
\footnote{This is not a necessary assumption. Smaller shocks like the ones in Finland and Britain discussed in the Section \ref{sec:empirical}  do not lead to negative $g_{11}$ but only to an exchange in the relation between the intrinsic growth rates of different parts of the economy: from $g_{11} > g_{22}$ to $g_{11} < g_{22}$. To detect with precision such shocks one needs finer GDP time resolution : quarterly rather than annually}.
The flows of capital between the sectors are parameterized by the off-diagonal  $\vec{G}$ elements. We will assume later that the regulators have the possibility to modify those flows.

The analytic solution given by Eq. \ref{eq:analytical} is shown in Fig. \ref{fig:crossing}. The total size of the economy is cf. Eq. \ref{eq:Ktot} represented by $K_{tot}(t)=k_1(t)+k_2(t)$ and has a 'crossing exponentials' shape with the old sector $k_1(t)$  initially decaying and the new sector $k_2(t)$  increasing. This illustrates the underlying Schumpeterian creative destruction economic process: the old sector decays exponentially while the new sector expands exponentially.

The empirical data of Fig. \ref{fig:Poland} clearly confirms this behavior: immediately after the shock the economic activity in some of the counties more than doubles while in the rest of the country it almost halves. 

This 'crossing exponentials' (Eq. \ref{eq:analytical}) behavior after a shock has been empirically confirmed in tens of cases \cite{Challet 2009}. A sharper version of the 'crossing exponential' effect happens when a new, second shock takes place while the system is still recovering from the first shock.  This sudden new switch between the leading and decaying sectors leads to the prediction of a very specific signal shown in Fig. \ref{fig:double}, which is very closely confirmed by empirical data Fig. \ref{fig:Balkan}: the initial sum of decaying+growing exponentials is abruptly broken and a new one, with a similar crossing exponentials structure, Eq. \ref{eq:analytical}. Consequently, a very sharp cusp signal is associated with the timing of the second shock.
The off-diagonal terms of  $\vec{G}$ are important because:
\begin{itemize}
\item  the initial resources of the old sector 1 are being used through $g_{21}>0$  to jump-start the new sector 2. (even an exponential factor cannot enhance a (almost) vanishing initial value).
\item Reciprocally, the exponentially growing wealth produced by the new sector 2 is eventually used through $g_{12}>0$  to support the growth of the otherwise decaying old sector 1. 
\item eventually after shrinking to its new natural size, the old sector will grow at the same rate with now leading new sector . Its relative size to it will be dictated by the flow of capital $g_{12}$ that it gets from the new sector.
\end{itemize}

More practically relevant, by controlling $g_{21}$  and  $g_{12}$, a government can enforce its policy by stimulating or suppressing the transfer of funds between the two sectors. Let us assume that initially the two sectors have been independent  $g_{12}=g_{21}=0$ and the government decides to use taxes to compensate partially their inequality \cite{Challet 2009}. The resulting $\vec{G}$  matrix will be: 
\begin{equation}
\vec{G}=
\begin{bmatrix}
g_{11}-\mu /2 & \mu /2 \\ \mu /2 & g_{22}- \mu /2 
\end{bmatrix}
\label{eq:G Fig 4}
\end{equation}
\begin{figure}
  \includegraphics[scale=0.45]{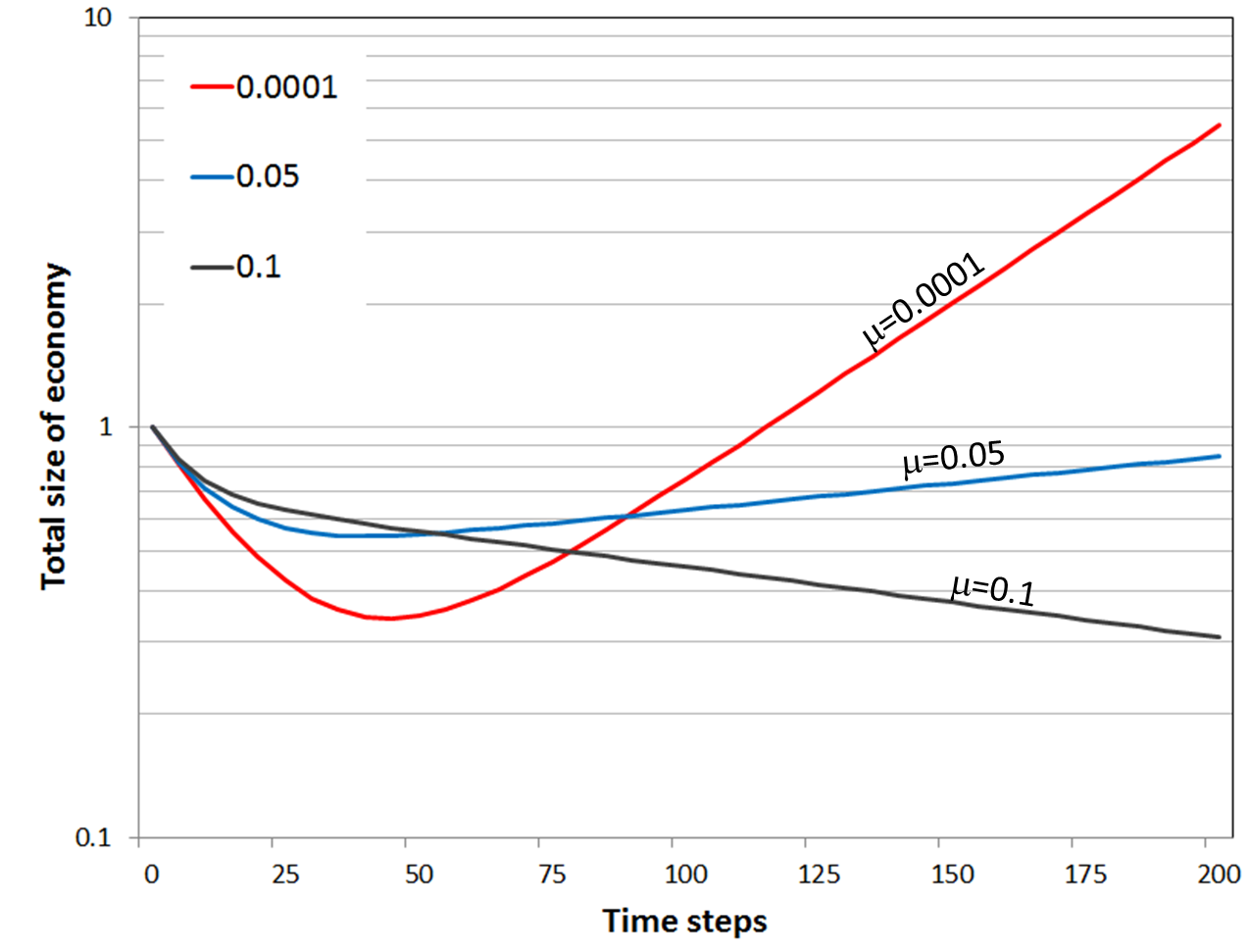}
\caption{{\bf Size of total economy versus time for various values of the transfer policy parameter $\mu$}
\\ The initial values are  $\vec{K}(0)=[k_1(0)=0.1, k_2(0)=0.9]$. 
\\ The 2x2 growth matrix after the shock is  
\\ $\vec{G}=[g_{11}=(0.02-\mu /2), g_{12}=  \mu /2,  g_{21}=  \mu /2,  g_{22}=(-0.05-  \mu/2)] $. 
\\ One sees there is a trade-off between avoiding short-term sacrifices and between ensuring long term prosperity:
\\-  while the hands-off shock-therapy in post-liberalization Poland was {\it a posteriori} criticized for its extreme human sufferance costs, it insured as seen in the Fig. \ref{fig:Poland} the quick recovery of even the least developed corners of the country. This is similar to the red $\mu = 0.00001$ graph. 
\\ - on the other hand, the Soviet reluctance to allow the creative destruction of the artificially upheld industrial and agricultural production modes let eventually to its demise. This is similar to the $\mu =0.1$ graph.
\\ - somewhat in the middle, the current policy of the US government and Fed has too many elements of supporting old dominant economic institutions and industries on the expense of the new emerging ones. One should be aware that this might affect the long term future growth as in the pink $\mu =0.05$ graph
(similarity to the Japan handling of their {\it hokai} crisis \url{http://en.wikipedia.org/wiki/Lost_Decade_(Japan)}).
}
\label{fig:Challet}       
\end{figure}

Fig. \ref{fig:Challet} shows the effects of the transfer policy for different values of the transfer parameter $\mu$. For its relevance to real economic crises see the discussion in Section \ref{sec:empirical}.  In the case when there is almost no transfer between sectors (red curve, $\mu=0.0001$), the recovery is very strong but only after having a very large and deep recession bottoming out at roughly $t = 45$.
By that time many companies and humans may have incurred irreparable damage and suffering.
Such a scenario is close to what actually happened during the shock therapy in Poland.

A moderate transfer can make the crisis shallower but longer in time (blue line, $\mu=0.05$).
Moreover, this would imply a longer recovery period and a penalty in the growth rate with which the economy exits the crisis. The example of Japan comes to ones mind.

 An exaggerated wish for equality, enforced by a large transfer could even lead to a continuous crisis and no recovery at all. This is seen in the $\mu = 0.1$ curve in Fig. \ref{fig:Challet}. The collapse of the communist economies is an illustration of this effect. 

 This effective analysis method allows for prediction of the effect of intervention policies and the speed of the recovery. Challet et al \cite{Challet 2009} considered a dynamically optimized transfer between sectors in which $\mu$ changes with the stages of the process. This allowed to 
reduce the initial shrinking /destruction of the economy due to the negative growth rate of the 'old' sector 
 without endangering the speedy and full recovery of the economy.   

Let us now discuss the behavior of the solution Eq. \ref{eq:analytical} for long periods of time between the jumps in $\vec{G}$. For $t \rightarrow \infty$  both components  $k_1(t)$ and  $k_2(t)$ of  $\vec{K}$   will grow exponentially with the same exponent equal to the highest eigenvalue $\lambda_{max}$:
\begin{equation}
\vec{K}(t \rightarrow \infty)=< \vec{K}(0), \vec{K}_{max}> e^{\lambda_{max} t} \vec{K}_{max}
\label{eq:Kmax}
\end{equation}
This leads to the conclusion that the growth rate of the total economy as well as the growth rate of the individual sectors will align with the slope given by the maximal eigenvalue $\lambda_{max}$. The relative size of the sectors will be then fixed in asymptotic time and equal to the ratio of the components of eigenvector $\vec{K}_{max}$:
\begin{equation}
\frac{k_1(t \rightarrow \infty)}{k_2(t \rightarrow \infty)}=\frac{k_{1max}}{k_{2max}}
\label{eq:sector ratio}
\end{equation}
This property that for asymptotically large times, in between $\vec{G}$ shocks, the economy reaches a steady state in which the sectors have the same exponential growth is true in general according to Frobenius-Perron theorem. One can see the empirical confirmation of this alignment of the growth rates in Fig. \ref{fig:Poland} after 1995. See the alignment of different sectors within a national economy as well as the alignment of different national economies in \cite{Dover 2009}

During the period of stability, the matrix  $\vec{G}$ has properties that recall very much the Leontief matrix. However, Leontief did not consider shocks in  $\vec{G}$ nor their implications for the growth rates: 
\begin{itemize}
\item the divergence of growth rates immediately after the shock, 
\item the crossing exponential in the transition period and 
\item the alignment of the growth rates for asymptotic times. 
\end{itemize}
Thus, one of the criticisms to Leontief theory was that different countries would have different growth rates. The above analysis shows on the contrary, that once the transfers $g_{ij}$  between the components (counties or countries) of an interacting economic system are introduced, the model predicts the asymptotic convergence of their growth rates. The theoretical analysis and the empirical confirmation of this effect in the inter-countries data was presented in detail in \cite{Dover 2009}. 

The present framework has a significant unifying potential in as much as it merges in a common 
conceptual framework the ideas of Schumpeter creative destruction, an agent model extension of the $AK$ model, Leontieff matrix,  Pareto scaling, Levy distributions, Simon-Mandelbrot skew distributions and cyclic economic behavior.  

\section{Empirical validation of the theoretical predictions }
\label{sec:empirical}
We discuss below some empirical results that confirm the theoretical analysis in the previous sections.  

\begin{figure}
  \includegraphics[scale=0.45]{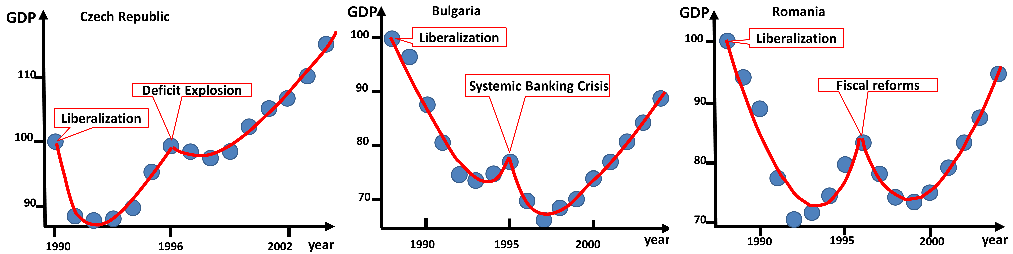}
\caption{Crossing exponentials events in real sector economy of different countries \cite{Challet 2009}: Bulgaria (a), Czech republic (b) and  Romania (c)  caused by different types of financial shocks. In these cases, it was possible  not only to pin-point the moments of the shock but in fact one could identify the actual events that caused those shocks. Some of those shocks were endogenous (deficit) and some exogenous (government and policy change). One sees that in confirmation of the main hypothesis of this article
(and in agreement with the Schumpeter / Minsky scenarios) the changes in the economic conditions can be described by discrete events where the $\vec{G}$ matrix changes. Indeed this leads to sharp breaks in the economic dynamics as predicted in Fig. \ref{fig:double}. In between those changes the system is quite precisely following the crossing exponential behavior  Eq. \ref{eq:analytical}. 
\\  Note that the graphs confirm a strong prediction of the mezo-economic Schumpeter matrix creative destruction model:
 the fluctuations can be decomposed into {\bf crossing exponentials ($J$ shape)} episodes with smooth minima separated by {\bf cusp ($\Lambda$ shape) maxima}.
Note this is opposed to the typical depiction of the economic cycle fluctuations where both the maxima and the minima are smooth. The pattern is universal and appears in many other autocatalytic systems in different disciplines.
}
\label{fig:Balkan}       
\end{figure}

The cross-exponentials recovery pattern was detected in many ex-communist countries after the fall of the old regime. In several countries, even a 'double cross-exponentials' pattern
(2 successive cross-exponential patterns) as depicted in Fig. \ref{fig:Balkan}) has been recorded. 

In a high resolution time series of Britain and Finland GDP 1990-2008, the entire range of 70 quarters was fitted with high precision in terms of cross-exponential episodes separated by cusps representing shocks / crises \cite{Challet 2009}. In the case of Finland one had just 2  cross exponential episodes: the first between the 1990 recession and the dot.com bubble burst in 2001 and the second between 2001 and the 2008 recession. In the case of Bitain one had an additional shock in 1995 connected to the real estate bubble burst and an additional small one in 2004.   
 
This demonstrates the possibility to precisely identify the shocks that start the economic cycles and model the economy as a sequence of such Schumpeterian creation-destruction episodes in terms of collective objects / herds that appear and disappear after shocks ({\bf marked by up-oriented cusps ($ \Lambda$ shape)}).
The identification of the granular structure of the economy in terms of those large lumps of capital / production allows the evaluation and prediction of the expected fluctuations in the 
economy as a whole.

The plots in Fig. \ref{fig:Balkan} start at the beginning of the liberalization shock. All three countries in Fig. \ref{fig:Balkan} underwent the second shock in the moment when they were in a positive growth phase. As mentioned in the Fig. \ref{fig:Balkan} the causes of the shocks were different: deficit, austerity program, banking crisis. Yet the 3 countries present the same pattern: the initial decaying exponential+growing exponential pattern was interrupted for a second one to take place.

 This creates the pattern which when iterated is the hallmark of all the autocatalytic models (microscopic and mezoscopic) described in this paper: fluctuations that consist of {\bf cusps pointing up ($\Lambda$ shapes) separated by smooth valleys ($J$ shapes)} corresponding to the crossing  exponentials  of the decaying old and expanding new dominant economy components. 

The models described in this paper may inform the optimal redistribution of resources policy within the economy. One can think how to transfer resources from the new emerging sectors to the old bulk of the economy or vice-versa. According to Fig. \ref{fig:Challet} the policy of transfers between the stronger and weaker sectors can make the crisis shorter but deeper, more violent and severe or the opposite.  The red line ($\mu=0.0001$), corresponding to minimal government intervention is somewhat similar to the case of Poland: the effect of the liberalization shock was very deep but very short and the recovery was very dramatic exiting the crisis with a very high growth rate. The blue line ($\mu=0.05$) in Fig. \ref{fig:Challet} is similar with the case of Japan, where upon the triggering of the crisis, the government decided to intervene maximally to soften  the crisis.  Contrary to the Schumpeterian doctrine of letting the crisis purge the economy 
of old, unsustainable production methods and technologies the Japanese government did not hesitate to transfer whatever amounts were necessary in order to preserve the existing businesses and sectors. As predicted by the Schumpeterian ideas, this made the crisis much milder but it prolonged it. 

Another possibility, corresponding to the black curve in Fig. \ref{fig:Challet} is to subsidize the old sector even beyond the minimum necessary to prevent its immediate economic collapse.  Paradoxically, the analysis of the matrix Schumpeter creative-destruction model  shows that this can be the most dangerous choice. An example of the results of such a policy was the case of Soviet Union. There, the government intervened continuously and systematically in order to subsidize the old obsolite economy on the expense of the sectors which -  if allowed - had a chance to produce growth. The result was that there was no recovery but going down from bad to worse. Our models showed that one can design an interactive policy in which one does not fix from the beginning the amount of transfer but does it by momentary optimization \cite{Challet 2009}. Then one can optimize the policy ensuring both the shallowest crisis and the fastest and most successful recovery. 

The transient regime after shock and in particular the growing + decaying (construction+destruction) crossing exponentials have been validated in tens of post-shock examples in different countries in \cite{Challet 2009}. The case studied in the finest detail was based on the annual data for the 3000 regions composing Poland in the years after liberalization \cite{Yaari 2008}. This was one of the cleanest examples of a creative destruction (of the legacy of communist economy) which not only the regulators did not try to harness but in fact deliberately initiated and allowed to unfold to its natural outcome. Because of this, the liberalization in Poland rightly deserved the label of shock therapy. The Balcerowicz program (as inspired by the ideas of Jeffrey Sachs) has led to a lot of suffering in the first years but it lead to a very good take off by the entire country very quickly. 

\begin{figure}
  \includegraphics[scale=0.45]{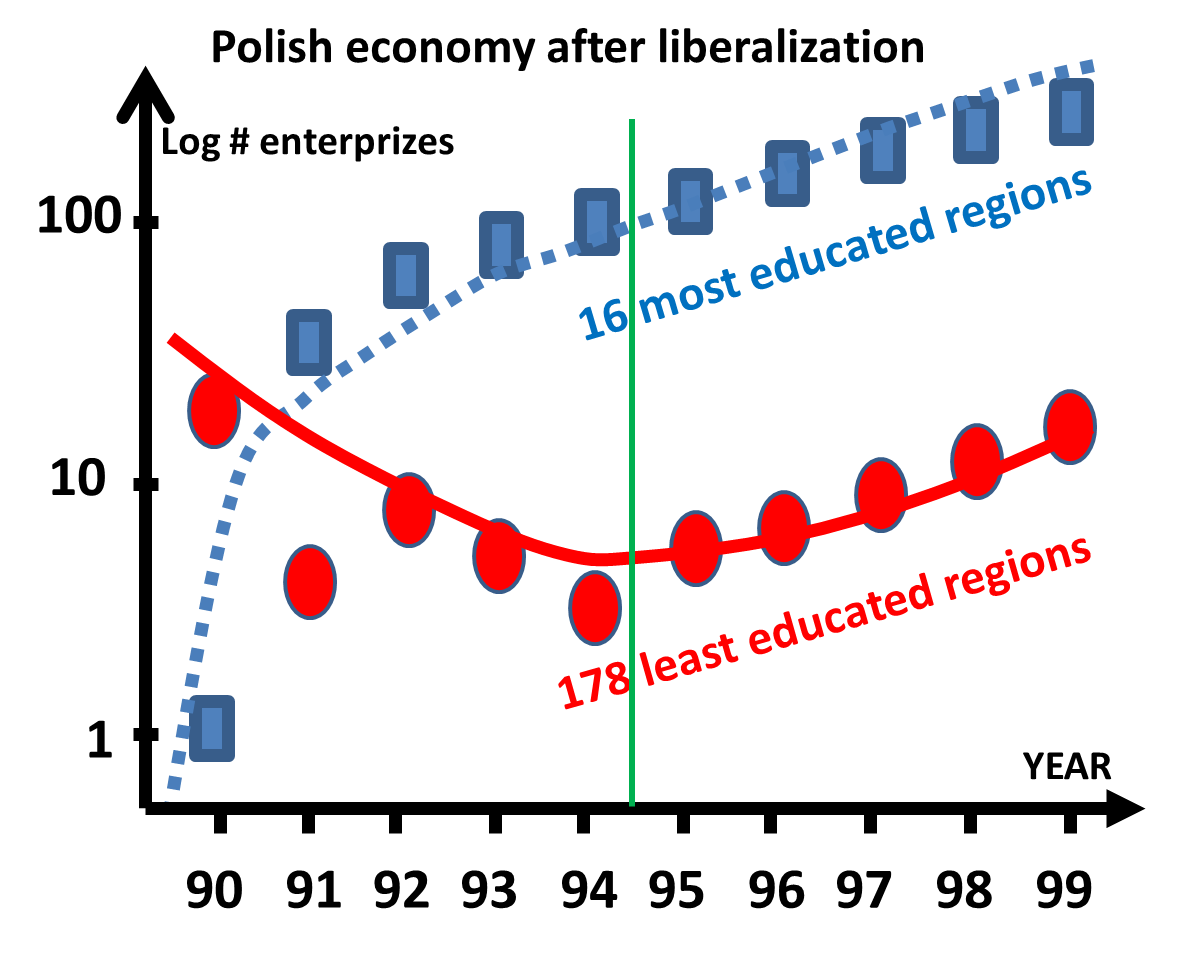}
\caption{The evolution of different components (gminas herds) of the Polish economy following  the liberalization.
\\  The blue squares are based on the empirical data in \cite{Yaari 2008} for the evolution of the most educated counties (average 12 years of schooling or more). 
\\ The red ovals are based the empirical data in \cite{Yaari 2008} for the least educated counties (average 8 years of schooling or less). 
\\  The (blue dotted and red continuous) lines represents a  2x2 matrix Schumpeterian 
numerical experiment as in Fig. \ref{fig:crossing}. The graphs and data were shifted vertically to account for the different number of counties in each set (16 for the most educated and 178 for the least educated). No effort was made to achieve exact fit between the real system and the 2x2 numerical experiment, but it is clear that the 2x2 captures the qualitative characteristics of the empirical data: 
\\ - the divergence of the growth rates during the initial Schumpeter creative destruction phase (before the vertical green line), 
\\ - the crossing exponentials, and 
\\ - the asymptotic Frobenius-Perron
alignment of the growth rates, Eqs. \ref{eq:Kmax}, \ref{eq:sector ratio} (after the vertical green line).
}
\label{fig:Poland}       
\end{figure}

The Poland post-liberalization scenario fits exactly the theoretical predictions of model Eq. \ref{eq:AKmatrix} and its agent based version Eqs. \ref{eq:s} and \ref{eq:delta}: Immediately after the shock a few growth centers (a herd of 16 counties represented by the black curve in Fig. \ref{fig:Poland}) emerged with growth rates of 400\% per year. At the same time, the economy of most of the country (the herd represented by the pink curve) collapsed by 50\%. Eventually, by diffusion, the economic recovery spread quickly from the initial herd of 16 counties throughout the entire economy. After only 5 years the system reached the steady state (the right side of the Fig. \ref{fig:Poland} after 1995) where, as predicted by the model, the various regions converged to a common growth rate. 

 Note that the convergence in terms of the growth rates is very clear while in terms of production per capita the data for different counties is still divergent:
 in Fig. \ref{fig:Poland} the difference between the number of enterprises per capita in the most (black) and the least (pink) educated counties continues to increase after 1995 exponentially with an exponent equal to their common growth rate.  
This explains why the convergence-divergence debate in economics was never convincingly settled \cite{Romer 1986}, \cite{Barro 1991}, \cite{Lee 1995}, \cite{Romer 1989}, \cite{Romer 1990}, \cite{Solow 1956}. The  various measures designed to detect convergence:  $\beta$-convergence, $\sigma$-convergence but were not able to find definitive evidence for it. This is because they looked for convergence in the absolute values per capita of the domestic product. According to the present (microscopic agent based and mezoscopic matrix-like) models (e.g. \cite{Yaari 2008}),  what is converging is not the domestic product per capita of the different regions, but rather the corresponding growth rates of it. Once it is realized that this is what should be measured and compared, it becomes evident e.g. Fig. \ref{fig:Poland} that such a convergence does take place. 

\section{Concluding remarks}
\label{sec:conclusion}
One of the important lessons from Masanao’s work is the correction of a widespread misunderstanding in the economics mainstream.
This misunderstanding produced and continues to inflict great damage to the advancement of economics. We  are referring to the urban legend that agent based modelling means simulations. While there are studies of agent based models that use simulations \cite{Delli Gatti 2008},  \cite{LLS 2000} and much of the results described above can be reproduced by massive Monte Carlo simulations, we intentionally emphasized the fact that with appropriate techniques, one can obtain important and illuminating analytical results. Those results guide the theoretical understanding of the system in terms of appropriate concepts: scaling, fractals, localization, singularities, phase transitions, non-ergodicity, etc.  
The multi-component generalization of the $AK$ model Eq. \ref{eq:AKmatrix}  capable to unify within the same 
theoretical formalism the short term fractal dynamics of markets Section \ref{sec:Levy walk} and the long term economic cycles Section \ref{sec:Predictions} is an indication of the progress one can be expected for the future.

The extraction of quantitative predictions from the multi-agent spatially extended stochastic models that express the non-equilibrium, non-self-averaging ideas of Masanao Aoki require results from field theory, master equation, percolation, branching random walks, stochastic differential systems, renormalization group which we tried to reproduce above in very simple terms. The present paper follows Aoki’s lead in meeting the great challenge to overcome the language and social barriers between the disciplinary domains and connect the physics and mathematics methods, ideas and techniques to the economic and financial domains that so badly need them. In the next phase, one has to make the results, conclusions and tools available to the practitioners and the decision makers in order to help them react to otherwise unexpected crises and opportunities. 

The autocatalytic feedback loops techniques touched upon in this paper are capable of transcending the usual linear causality and to express in rigorous terms the spontaneous emergence of qualitatively novel properties as the collective effect of a multitude of random causes. In order to understand and control the evolution of mass phenomena such as the current crisis, one has to give up the concept of linear causality chains which associate to each effect a cause. One has to think, rather, in terms of a collective causality emerging from the microstructure. The existence of interactions between the components that constitute the system and the system as such renders the system capable of endogenous creativity and innovation.  

The spontaneous emergence of self-organized adaptive collective objects is the key to a functional economy. With all the penalties that that free capitalist economy has to pay for its heterogeneity and inequality and for its fractal and cyclic intrinsic fluctuations, this seems the only viable alternative.
As argued by Schumpeter and Minsky, any attempt to eliminate wealth heterogeneity and / or 
time fluctuations in the economy, renders the system dysfunctional. This was confirmed empirically along the history through the demise of so many regimes that tried to ignore it.

This understanding opens the way of bringing the economic models closer to the reality as perceived by Masanao: capable of endogenous innovation, capable of transcending the mere implications of their previous time evolution, capable to generate collective objects with properties that are completely different from the intentions, scales and scope of the original individual components \cite{Yaari 2008}, \cite{Cantono 2010}, \cite{Solomon 2013}.

As emphasized by Masanao in his work on non-self-averaging \cite{2}, \cite{3}, \cite{Aoki 2008}, a crucial aspect of economic systems is the irrelevance of statistical averages: while one can define an average over different realizations of the system, this will be often irrelevant for each individual realization: once specific individual rare events are promoted to systemic changes, they cause the system to take dramatically different time trajectories. 

Masano’s singular path in science may in itself be an example of an instance in which a very rare event influenced the collective macroscopic dynamics of the entire system.

\appendix
\section*{Appendix \\ The connection between intermittent / fractal / macroscopic fluctuations, and the anomalous scaling of the central peak $P_{origin} (t)$ }

Eq. \ref{eq:Polyasum} suggests $P_{origin} (t)$ as a generic measure of the stability of a system under random fluctuations: if $P_{origin} (t)$ decays slower than $1/t$ then the fluctuations are eventually bringing the system back to its original value. If on the contrary $P_{origin} (t)$ decays faster then $1/t$
the fluctuations will eventually bring the system arbitrarily far away from the original value.

Moreover, we will see that the behavior of $P_{origin} (t)$ informs about the nature of the individual steps
of which the random walk consists. Since the size of the individual steps are related to the individual collective objects that compose the system, $P_{origin} (t)$ can inform about the granular structure of the system. It is interesting that the time fluctuations of the global measurables are revealing the actual composition of the system.  For instance the fluctuations of $K(t)$ can inform of the collective objects (herds, sectors, capital aggregates)  $K_i (t)$ composing it:
\begin{equation}
K(t) =  K_1(t)+ K_2(t) + ... +K_N (t)
\label{eq:K sum}
\end{equation}
We will use this in order to characterize economic systems in various regimes:
\begin{enumerate}
\item if the inhomogeneous parts $K_i$ of $K(t)$  (islands, herds, mountains) are small and mainly limited to the micro-scale, then their appearance, disappearance and motion amount to a random walk of the total $K(t)$  with limited sized steps $\Delta K(t)$. Such 'microscopic' steps will lead to a Gaussian, normal random walk whose fluctuations are negligible at the macroscopic level.
\item If the dynamic collective objects composing $K(t)$ are scaling, e.g. if their sizes are distributed by a Pareto-Zipf power law 
\begin{equation}
K_i \sim i^{-1/\alpha}
\label{eq:K Zipf}
\end{equation}
  then the $K(t)$ fluctuations are scaling as 
\begin{equation}
P_{origin} ( t) \sim 1/\sigma(t) \sim t^{-1 / \alpha}. 
\label{eq:Origin scaling}
\end{equation}

Typically for the individuals' capital in the western economies $1<\alpha <2$ which insures the same $\alpha$ for the stock market index $P_{origin}(t)$. See Fig. \ref{fig:Pareto} for empirical realizations of this prediction.
 For $\alpha > 2$ one returns to the regime 1.
\item For $\alpha <1$ the majority of the capital $K(t)$ is concentrated in the first terms in Eq.  \ref{eq:K Zipf} (e.g. dominant economic sectors $K_1$, $K_2$).
Thus the fluctuations themselves follow the macroscopic changes in those $K_1$, $K_2$ parts and in particular in their crossing: the substitution - following a shock - of the largest $K_1 (t)$ by a new one.
The result is intermittent dynamics of the economic cycles type.
However as opposed to the typical depiction of the economic cycle fluctuations where both the maxima and the minima are smooth, in our case {\bf maxima are cusps} and in between $K(t)$ is a smooth sum of 2 exponentials (corresponding to the old and new dominant $K_1(t)$'s) (e.g. Fig. \ref{fig:Balkan}). 
\end{enumerate}

\subsection*{Collective objects and time changes with limited size}

Let us start with the case when the step sizes have a flat probability density uniformly distributed between $-D$ and $D$:
\begin{equation}
 P ( \Delta X(t) )  = \begin{cases}
    1/(2D), & \Delta X(t) \in [-D,D]\\
    0, & \text{otherwise}.
  \end{cases}
\label{eq:gauss term prob}
\end{equation}
For long time intervals, during which many individual steps $t$ took place, the accumulated change in $K(t)$ is the sum of the individuals steps:
\begin{equation}
 \Delta K(t) = \sum_t \Delta X(t)
\label{eq:gauss sum}
\end{equation}
Consequently  the variation of $K(t)$ within a time interval of $t$ individual steps 
is given according to the central limit theorem by a Gaussian, normal distribution:
\begin{equation}
 P_{Gauss} (\Delta K,t) ={4 \pi D t}^{-1/2} e^{- (\Delta K)^2 / (4 D t)},
\label{eq:gauss}
\end{equation}
 with a  standard deviation extending as $\sqrt {t}$:
\begin{equation}
\sigma_{Gauss} ( t) =   2 \sqrt{ D t}
\label{eq:sd}
\end{equation}

The peak of this distribution Eq.\ref{eq:gauss} decays for long time intervals $t$ like a power $-1/2$ of $t$:
\begin{equation}
 P_{origin Gauss} (t) = P_{Gauss} (\Delta K=0,t) ={4 \pi D t}^{-1/2}.
\label{eq:gauss height}
\end{equation}
This is a general property of the time evolution of probability stable distributions
(distributions that preserve up to a scale their shape):
since their integral has to remain $1$,
the increase of  their width in time has to be compensated by a corresponding lowering in time of  their height: 
\begin{equation}
P (\Delta K=0,t)= P_{origin} (t) \sim 1/ {\sigma} (t)
\label{eq:gauss width}
\end{equation}

Reciprocally since the integral of the probability density over all the possible values of $\Delta K$  has to be $1$, a slow decrease like \label{eq:gausspolya} in its height at the $\Delta K(t)=0$ peak  $P (\Delta K=0,t)$ implies a slow expansion of its width with roughly the same rate: 
\begin{equation}
{\sigma}(t) \sim 1/ P (\Delta K=0,t) = 1/ P_{origin} (t)
\label{eq:gauss sigma}
\end{equation}
Thus a for slow decays of $P_{origin} (t)$ with $t$ one will have an evolution of $K(t)$ for which the stochastic part
 $ \sigma (t) \sim {D t}^{1/2}$ 
does not  take it far away from the causal part which is typically $\sim t$.

This is important for the study of the properties of the $K(t)$ fluctuations.
Indeed, by definition the extreme fluctuations are exceedingly rare and influenced by the finite size of the system and of the observation time. This makes the measuring of the 
extreme fluctuations nearly impossible and also useless from the theoretical point of view (because of the contamination of the 'ideal' 'theoretical' values by finite size and finite time corrections. 
Thus connecting the large fluctuations behavior of $K(t)$ to the 
behavior of the central peak is an important step to obtain information on the character of the 
fluctuations of $K(t)$ and to discriminate between the various regimes with great reliability and precision:
\begin{itemize}
\item As opposed to extremely large and extremely rare events with $\Delta K >> \sigma$ the frequency of events with $\Delta K \sim 0$ is very high, insuring high statistics, high precision measurements. 
\item The small size (in fact $0$) of the variation $\Delta K =0$ makes the measurement less sensitive to the finite size and finite time effects. Of course indirectly, the finite size enters in the statistics of the ensemble of measurements values but it does not affect directly and strongly in the actual $\Delta K =0$  measured value of the events entering $P_{origin} (t)$ (as it does for measurements of $P(\Delta K,t)$ with very large $\Delta K >> \sigma$ \cite{Mantegna 1999}). 
\end{itemize} 

Consequently we used
\cite{Malcai 1999}, \cite{Solomon 2001}, \cite{Solomon 2003}, \cite{Klass 2007}  
 the time decay of the central peak $P_{origin} (t)$ of the probability density of $ P(\Delta K,t)$ as a measure of the stability of the dynamics of $K(t)$. I.e. of the behavior of its large, fractal fluctuations.

\subsection*{Collective objects and time changes that reach to larger scales}

By contrast to the case in which the size of the individual steps is bounded to a microscopic scale $D$
 the  fluctuations of the   'herds' / 'islands' / 'large fortune' 's sizes in the autocatalytic models presented in this paper may spread up to the macroscopic / system scale.

Assume the objects $i$ composing the system of size $K(t)$ are ordered in decreasing size $K_i (t)$ order.
Assume they follow at each moment a Pareto-Zipf law:
\begin{equation}
K_i \sim i^{1/\alpha}  
\label{eq:Zipf law}
\end{equation}
(not necessarily in the same order $i$ at different times).

Equivalently the probability for a collective object to have a size more then some value $K_{collective \ object \ size}$ is
\begin{equation}
P_{Pareto} (K_i> K_{collective \ object \ size})  = ({K_{collective \ object \ size}})^{-\alpha} ; 
\label{eq:ParetoZipf}
\end{equation}

Thus the appearance / disappearance / shrinking / growth of the collective objects  consists from the point of view of $K(t)$  into a fractal / intermittent random walk. 
 I.e a random walk in which the step sizes $\Delta X(t)$ have a scaling probability distribution:
\begin{equation}
P_{Pareto} (\Delta X > \Delta Z) = {\Delta Z}^{-\alpha} ; 
\label{eq:PstepLevy}
\end{equation}

The properties of such random walks were introduced by Paul Levy and popularized by Mandelbrot \cite{Mandelbrot 1982}. Instead of a Gaussian process, the resulting $\Delta K(t)$ dynamics
 will consist of a Levy flights  process $L_{\alpha}(\Delta K,t)$ of index $\alpha$.

There is no analytic formula for the Levy probability density $L_{\alpha}(\Delta K,t)$.
But one can deduce the time evolution of its  width $\sigma_{Levy} (t)$ and of the height of its central peak $P_{Levy \ origin} (t)$ (recall their product is a constant because of the integral of a probability density has to remain 1 at all times)  from the following argument.

According to Eq. \ref{eq:PstepLevy} the waiting time for a step of size at least $\Delta Z$ to appear is : 
\begin{equation}
t(\Delta Z) =(\Delta Z)^{\alpha}; 
\label{eq:TstepLevy}
\end{equation} 
Thus, reciprocally the largest value of a single step $\Delta Z$ expected to appear within a time interval $t$ will be
\begin{equation}
\Delta Z \sim t ^{1/\alpha}; 
\label{eq:LevyTstep}
\end{equation} 
For $\alpha < 2$ this single step dominates the contributions of all other steps to the variation of $K(t)$:
\begin{equation}
\Delta K(t)  = \sum_t \Delta Z(t) \sim \Delta Z_{max} (t) \sim t ^{1/\alpha}; 
\label{eq:Levy sum}
\end{equation}
Thus, the largest individual step expected to appear in a given time steps interval $t$ dominates the time dependence of the standard deviation $\sigma_{Levy} (t)$  of the probability density and consequently also its height $P_{Levy} (\Delta K=0,t)$:
\begin{equation}
P_{Levy \ origin} (t) \equiv P_{Levy} (\Delta K=0,t) \sim 1/ {\sigma}_{Levy} (t) \sim { t}^{-1/\alpha}.
\label{eq:Levy width}
\end{equation}
This means that the probabilityfor $K(t)$ to be back  at the  original value after a time $t$ decreases with $t$ as ${ t}^{-1/\alpha}$. 
This is faster than in the Gaussian case because if ${1/\alpha} >{1/2}$.
 Reciprocally, the speed by which the system diffuses away  from an original value that it had at the beginning of a time interval $t$ is ${\sigma}_{Levy} (t) \sim { t}^{1/\alpha}$,  faster than in the Gaussian case.
Moreover, both time evolutions are dictated by the same exponent $\alpha$ that characterizes the step size distribution Eq. \ref{eq:PstepLevy}. In turn the step sizes depend on the sizes of the collective objects that define the granularity of the system (in the present case on the Pareto-Zipf exponent $\alpha$ of the power law: Eq. \ref{eq:Zipf law}). This connection between the sizes of the mezo-economic objects (large wealths, geographical regions capital, economic sectors) and the time changes in the economic indices (market fractal fluctuations, GDP intermittent cycles) is the main message of our analysis.

The above phenomena even though exist at different time scales and fluctuations amplitudes are all 
the expressions of the same conceptual framework for different  $\alpha $ regimes:  
\begin{itemize}
\item For a granularity dominated by small collective objects $\alpha >2$ the approximation Eq. \ref{eq:Levy sum} does not hold because the finite / microscopic size steps \ref{eq:gauss term prob} already imply a faster spread Eq. \ref{eq:gauss sigma} of $\Delta K(t) \sim t^{1/2}$ cf. Eq. \ref{eq:sd}.
\item For a granularity dominated by the largest collective objects $\alpha <1$ the evolution of $K(t)$ due to fluctuations is faster than what would be expected from a regular causal dynamics with finite speed (i.e. change in $K(t)$ proportional to the time interval $t$).  in fact this means that the fluctuations are taking the values of $K(t)$ in an accelerated way far from its original position. Moreover, $\Delta Z_{max} (t)$ reaches very quickly the limits imposed by the finite size in the system implying that the dynamics reduces to steps that are of the system size: $K(t)$ can take "any value"; for long times, the system doesn't even have an average $ < K(t) >$.
\end{itemize}

In conclusion, depending on the level of granularity of the herds / sectors / capital accumulations  / production clusters involved we have 3 regimes of the time fluctuations of the global economic indices:

\begin{itemize}
\item For relatively small size clusters (limited scale steps or very rare  medium scale ones) {\bf $\alpha > 2$ } one recovers the Gaussian regime \ref{eq:gauss width}.
The time fluctuations remain microscopic and of negligible relative scale $(size \ of \ the \ system)^{-1/2}$.
\item For larges clusters / herds imposing larger unit steps that behave like Eq. \ref{eq:PstepLevy} with $1< \alpha <2$ the time fluctuations are not normal / Gaussian and are in the Levy / fractal domain in Mandelbrot terminology. This is the case of the stock market indices with the caveat that for very large values the limitations in the size of the system / its components affect the very distant tails \cite{Mantegna 1999}.
The connection between the granularity of the individual wealth (measured by the Pareto exponent in Eq. \ref{eq:LevyTstep}) and the fractal exponent of the market index fluctuations (measured by the exponent of $P_{Levy \ origin} (t)$ Eq. \ref{eq:Levy width}) has been verified empirically in a number of countries \cite{Solomon 2003}. See Fig. \ref{fig:Pareto}.  
\item for a distribution in which the largest clusters are of the systems size (e.g. size steps scaling with $\alpha <1$ Eq. \ref{eq:PstepLevy}) the fluctuations are so strong that essentially the system is completely dominated by the largest steps sizes. We will see (Sections  \ref{sec:Predictions} and \ref{sec:empirical} \cite{Malcai 1999}, \cite{Solomon 2001} \cite{Solomon 2003}) that this establishes a very direct connection between the intermittent cyclic variations in $K(t)$ and the individual wealth / cluster / herd granularity .
\end{itemize}

\begin{thebibliography}{}
\bibitem{Anderson 1972} Anderson, Phil W. "More is different". Science, 177(4047):393-396, 1972.
\bibitem{Popper 1976} Popper Karl. (1976). "Unended Quest: An Intellectual Autobiography", Open Court, LaSalle, Illinois.
\bibitem{Schumpeter 1936} Schumpeter, Joseph A. "The theory of economic development". Harvard University Press, Cambridge MA, 1936.
\bibitem{Montroll 1978} Montroll, Elliott W. “Social dynamics and the quantifying of social forces.” Proceedings of the National Academy of Sciences 75, no. 10 (1978): 4633-4637.
\bibitem{Aoki 1968} Aoki, Masanao. “Control of large-scale dynamic systems by aggregation.” Automatic
\bibitem{Romer 1986} Romer, Paul M. "Increasing returns and long-run growth." The Journal of Political Control, IEEE Transactions on 13, no. 3 (1968): 246-253.  Economy (1986): 1002-1037.
\bibitem{Mandelbrot 1982} Mandelbrot, Benoit B. "The Fractal Geometry of Nature" (Updated and augm. ed.). New York: W. H. Freeman (1982).
\bibitem{Zeldovich 1987} Ya. B. Zel'dovich, S. A. Molchanov, A. A. Ruzmaikin, and D. D. Sokolov "Intermittency in random media" Usp. Fiz. Nauk 152, 3-32 (1987).
\bibitem{Kirman 1992} Kirman, A.“What or whom does the representative individual represent?”, Journal of Economic Perspectives, 6(2), 117-36.
\bibitem{von Hayek 1937} Von Hayek, Friedrich A. "Economics and knowledge." Economica (1937): 33-54.
\bibitem{Aoki 2004} Aoki, Masanao. "New frameworks for macroeconomic modelling: some illustrative examples." UCLA Economics Online Papers (2004).
\bibitem{Malcai 1999} Malcai Ofer, Biham Ofer and Solomon Sorin. "Power-law distributions and Levy-stable intermittent fluctuations in stochastic systems of many autocatalytic elements". Physical Review E, 60, 1299-1305 (1999).
\bibitem{Solomon 2003} Solomon, Sorin, and Moshe Levy. "Pioneers on a new continent: on physics and economics." Quantitative Finance 3, no. 1 (2003): c12-c15.
\bibitem{Klass 2006} Klass, Oren S., Ofer Biham, Moshe Levy, Ofer Malcai, and Sorin Solomon. "The Forbes 400 and the Pareto wealth distribution". Economics Letters 90, no. 2, pp. 290-295 (2006).
\bibitem{Klass 2007}  Klass, Oren S., Ofer Biham, Moshe Levy, Ofer Malcai, and Sorin Solomon. "The Forbes 400, the Pareto power-law and efficient markets." The European Physical Journal B 55, no. 2 (2007): 143-147.
\bibitem{Gabaix 2011}  Gabaix Xavier "The Granular Origins of Aggregate Fluctuations," Econometrica,  vol. 79, p.733-772 (2011)
\bibitem{Dosi 2012} Dosi, Giovanni. "Economic Coordination and Dynamics: Some Elements of an Alternative 'Evolutionary' Paradigm." Institute for New Economic Thinking (downloaded from http://ineteconomics. org/blog/inet/giovanni-dosi-response-john-kay-elements-evolutionary-paradigm on 3 February 2012) (2011).
\bibitem{Yaari 2008} Yaari, Gur, Andrzej Nowak, Kamil Rakocy, and Sorin Solomon. "Microscopic study reveals the singular origins of growth." The European Physical Journal B-Condensed Matter and Complex Systems 62, no. 4 (2008): 505-513. 
\bibitem{Challet 2009} Challet, Damien, Sorin Solomon, and Gur Yaari. "The universal shape of eco-nomic recession and recovery after a shock." Economics: The Open-Access, Open-Assessment E-Journal 3 (2009).
\bibitem{Dover 2009} Dover, Yaniv, Sonia Moulet, Sorin Solomon, and Gur Yaari. "Do all economies grow equally fast?" Risk and Decision Analysis 1, no. 3 (2009): 171-185.,  
\bibitem{Malthus 1798} Malthus, Thomas. "An Essay on the Principle of Population As It Affects the Future Improvement of Society, with Remarks on the Speculations of Mr. Godwin, M. Condorcet, and Other Writers 1798". Publisher/Edition London: J. Johnson, in St. Paul's Church-yard.
\bibitem{Shiller 2000}  Shiller Robert J , Irrational Exuberance, Princeton University Press (2000)
\bibitem{Aghion 2008} Aghion Philippe,  Peter W. Howitt, "The Economics of Growth" The MIT Press (1997).
\bibitem{Verhulst 1838} Verhulst, Pierre-François, "Notice sur la loi que la population poursuit dans son accroissement". Correspondance mathématique et physique 10: 113–121 (1838).
\bibitem{Shnerb 2000} Shnerb, Nadav M., Yoram Louzoun, Eldad Bettelheim, and Sorin Solomon. "The importance of being discrete: Life always wins on the surface." Proceedings of the National Academy of Sciences 97, no. 19 (2000): 10322-10324.
\bibitem{Kesten 2002} Kesten, Harry, and Vladas Sidoravicius. "Branching random walk with catalysts." Electron. J. Probab 8 (2003): 1-51.
\bibitem{Louzoun 2007} Louzoun, Yoram, Nadav M. Shnerb, and Sorin Solomon. "Microscopic noise, adaptation and survival in hostile environments." The European Physical Journal B 56, no. 2 (2007): 141-148.
\bibitem{Vieira 2013} Santos, R. V. D. (2013). The importance of being discrete in sex. arXiv preprint arXiv:1305.0446.
\bibitem{Louzoun 2003a} Louzoun, Yoram, Sorin Solomon, Henri Atlan, and Irun R. Cohen. "Proliferation and competition in discrete biological systems." Bulletin of mathematical biology 65, no. 3 (2003): 375-396.
\bibitem{Louzoun 2003} Louzoun, Yoram, Sorin Solomon, Jacob Goldenberg, and David Mazursky. "World-size global markets lead to economic instability." Artificial life 9, no. 4 (2003): 357-370.
\bibitem{Goldenberg 2004} Goldenberg, Jacob, Barak Libai, Yoram Louzoun, David Mazursky, and Sorin Solomon. "Inevitably reborn: the reawakening of extinct innovations." Technological Forecasting and Social Change 71, no. 9 (2004): 881-896.
\bibitem{Huang 2001} Huang, Zhi-Feng, and Sorin Solomon. "Finite market size as a source of extreme wealth inequality and market instability." Physica A: Statistical Mechanics and its Applications 294, no. 3 (2001): 503-513.
\bibitem{Solomon 2001} Solomon, Sorin, and Peter Richmond. "Power laws of wealth, market order volumes and market returns." Physica A: Statistical Mechanics and its Applications 299, no. 1 (2001): 188-197.
\bibitem{Mantegna 1999}Mantegna Rosario N. and H. Eugene Stanley, "Introduction to Econophysics: Correlations and Complexity in Finance" Cambridge University Press (1999)
\bibitem{Mankiw 2006} Mankiw, Gregory N. (2006). "Principles of Economics", 4th Edition. South-Western College Pub. ISBN 0-324-22472-9.
\bibitem{Hidalgo 2007} Hidalgo, César A., Bailey Klinger, A-L. Barabási, and Ricardo Hausmann. "The product space conditions the development of nations." Science 317, no. 5837 (2007): 482-487.
\bibitem{Leontief 1947} Leontief, Wassily. "Introduction to a theory of the internal structure of functional relationships." Econometrica, Journal of the Econometric Society (1947): 361-373.
\bibitem{Aoki 1999} Aoki, Masanao, and Hiroshi Yoshikawa. "Demand creation and economic growth." U. of Tokio, Ctr (1999).
\bibitem{Barro 1991} Barro, Robert J.  and Sala-i-Martin, Xavier. "Convergence across states and regions". Economic Growth Center, Papers on Economic activity, (1991) 1:107-158.
\bibitem{Lee 1995} Lee, Kevin, M. Hashem Pesaran, and Ron P. Smith. "Growth and convergence in a multi-country empirical stochastic Solow model." Journal of applied Econometrics 12, no. 4 (1997): 357-392.
\bibitem{Romer 1989} Romer, Paul M. "Endogenous technological change". No. w3210. National Bureau of Economic Research, working paper,  December 1989.
\bibitem{Romer 1990} Romer, Paul M. "Endogenous technological change". Journal of Political Economy, 98(5):S71-S102, 1990.
\bibitem{Solow 1956} Solow, Robert M. "A contribution to the theory of economic growth." The quarterly journal of economics 70, no. 1 (1956): 65-94.
\bibitem{Delli Gatti 2008} Gatti, Domenico, Edoardo Gaffeo, Mauro Gallegati, Gianfranco Giulioni, and Antonio Palestrini. "Emergent macroeconomics: an agent-based approach to business fluctuations". Springer, 2008.
\bibitem{LLS 2000} Levy, Haim, Moshe Levy, and Sorin Solomon. "Microscopic simulation of financial markets: from investor behavior to market phenomena". Academic Press (2000).
\bibitem{Cantono 2010} Cantono, Simona, and Sorin Solomon. "When the collective acts on its components: economic crisis autocatalytic percolation." New Journal of Physics 12, no. 7 (2010): 075038.
\bibitem{Solomon 2013} Solomon, Sorin, and Natasa Golo. "Minsky Financial Instability, Interscale Feedback, Percolation and Marshall–Walras Disequilibrium." Accounting, Economics and Law 3, no. 3 (2013): 167-260.
\bibitem{2} Aoki, Masanao and R. J. Hawkins. "Non-self-averaging and the statistical mechanics of endogenous macroeconomic fluctuations". Economic Modelling, 27:1543-1546, 2010.
\bibitem{3} Aoki, Masanao and Hiroshi Yoshikawa. "Non-self-averaging in macroeconomic models: a criticism of modern micro-founded macroeconomics". Journal of Economic Interaction and Coordination, 7:1-22, 2012.
\bibitem{Aoki 2008} Aoki, Masanao, and Hiroshi Yoshikawa. "The nature of equilibrium in macroeconomics: a critique of equilibrium search theory." Economics: The Open-Access, Open-Assessment E-Journal 3 (2009).
\end{thebibliography}
\end{document}